\documentclass[%
 preprint,
superscriptaddress,
 amsmath,amssymb,
 aps, physrev,
]{revtex4-2}

\newcommand{\norm}[1]{\left\|#1\right\|}

\usepackage{amsthm}

\newtheorem{theorem}{Theorem}

\newtheorem{assumption}{Assumption}

\theoremstyle{definition}
\newtheorem{definition}{Definition}

\newtheorem{remark}{Remark}

\usepackage{graphicx}% Include figure files
\usepackage{dcolumn}% Align table columns on decimal point
\usepackage{bm}% bold math
\begin{document}

%\preprint{APS/123-QED}

\title{Dynamic models with $p$ parameters are identified by $2p+1$ random features}

\author{Michael Wieck-Sosa}
\affiliation{Department of Statistics \& Data Science, Carnegie Mellon University, Pittsburgh, PA 15213 USA}
\author{Cosma Rohilla Shalizi}
\affiliation{Department of Statistics \& Data Science, Carnegie Mellon University, Pittsburgh, PA 15213 USA}
\affiliation{Santa Fe Institute, 1399 Hyde Park Road, Santa Fe, NM 87501 USA}
 \email{Contact author: cshalizi@cmu.edu}

\date{\today}% It is always \today, today,

\begin{abstract}
A foundational principle in nonlinear dynamics is that the structure of a dynamical system can be recovered from a small number of generic measurements or coordinates. We develop an analogous principle for the identification of dynamic models for time series \textit{with noise}, which builds on previous identification results for noiseless dynamical systems. The noise is allowed to be non-iid, non-Gaussian, and dependent on the state. Our results cover noisily observed differential equations and discrete-time dynamical systems, as well as stochastic models with process noise. We illustrate the utility of this identification principle using a Lorenz-63 model and a H\'{e}non map model, both with observational noise. 
%\begin{description}
%\end{description}
\end{abstract}

\maketitle

\section{Introduction}

Since \citet{geom_ts_Packard_et_al_1980}, one aim of nonlinear dynamics has
been to recover the attractors of deterministic dynamical systems from time
series of a single observable function of the state.  Long before that, time
series analysts had practiced predicting the next observation from a
vector of delayed values of the observable, and it was intuitively plausible
that adding more delayed values would give more information about the state,
and so about future observations.  What made ``geometry from a time series'' or
``state space reconstruction'' into both a research program and a practical
tool was the proof by \citet{embedding_takens_1981} that, when the state had a
finite dimension, it could be reconstructed exactly, up to a smooth change of
coordinates, by a {\em finite} number of delayed observations, at most $2d+1$
when the state has $d$ dimensions.  Or, rather, all of this is true for
``typical'' or ``generic'' choices of the observable and delay.

``Geometry from a time series'' only allows one to reconstruct the unobserved
state up to a smooth change of coordinates.  As such, it cannot support
inferences about specific functional forms for the underlying dynamics, or for
numerical parameters describing those dynamics.  On the other hand, scientists
increasingly express their dynamical ideas as models with very specific
functional forms, with a variety of physically-meaningful but {\em a priori}
unknown parameters.  For the models to be useful, those parameters must be
estimated from data. Note that, among statisticians, inferring values of parameters from data is
called ``estimation'', while, e.g., climatologists are more
apt to speak of ``calibration''.  We will generally follow conventions of
statistics, but we will define terms clearly.

A logically prior question to estimation, however, is whether the parameters of
the model {\em can} be inferred from the data available, i.e., whether the
observed data contain any information about the parameters at all.  This is
what statisticians call ``parameter identification'', or just
``identification''. By contrast, what control engineering and signal
processing calls ``system identification'' is closer to ``estimation'' in statistics. Speaking loosely (we are more precise below), the parameter
vector $\theta$ is identified by the random vector $X$ when differences in
$\theta$ imply differences in the distribution of the random variable:
$\theta_1 \neq \theta_2$ only if $p(x;\theta_1) \neq p(x;\theta_2)$ for at
least some $x$.

When a model is {\em un-}identified, there are at least two
parameter vectors which lead to exactly the same distribution, hence no
conceivable data analysis can distinguish those parameters.  Most computational
models advanced in scientific practice would be identified, if we could observe
the full state of the system; we should, of course, be so lucky. There are rare exceptions with genuine
redundancies or symmetries in the model parameterization, but these are usually
accidental, and can be eliminated by, e.g., picking a canonical value from each congruence class.

Our main result here is two-fold.  First, for a large class of dynamical
systems observed through stochastic noise, a $p$-dimensional vector of
parameters can be identified through the expectations of $2p+1$ generic
functions of observables, or ``features''. In statistical theory, any
(measurable) function of the data is called a ``statistic''.  We find this term
leads to confusion among non-statisticians, so we will speak of ``features'',
as is common in data mining and machine learning. We may actually choose the features
{\em randomly}, from a fixed, data-independent distribution over a
suitably-rich class of basis functions, e.g., Fourier basis functions for
vector-valued processes. We show all this by adapting results from \citet{amir_et_al_finite_witness_thm}. Second, because the full distribution of those observables is not
required for identification, we can construct practical estimation procedures
for inferring the parameters from time series data, either in the long-time
(``ergodic'') limit for stationary or asymptotically-mean-stationary processes,
or in the dense-sampling (``infill'') limit for nonstationary processes.

\subsection{Related work}

As mentioned, the ``geometry from a time series'' program traces back to
\citet{geom_ts_Packard_et_al_1980}.  This initiated the procedure of forming
vectors from finitely many delayed values of a scalar signal, producing what is
now called a \textit{time-delay embedding}, and analyzing the resulting
time-delay embedding space.  \citet{embedding_takens_1981} established that using $2d+1$ delays yields a time-delay embedding space that is topologically equivalent to the original $d$-dimensional state space. The number of required delays can be smaller in many cases, with sharper bounds depending on the box-counting dimension of the attractor~\citep{embedology_Sauer_et_al_1991}. Time-delay embeddings became a major source of inspiration for later works in nonlinear dynamics; see~\citet{nonlinear_time_series_analysis_kantz_schreiber_2003} for a comprehensive overview.

The~\citet{embedding_takens_1981} embedding theorem has been extended in several directions.~\citet{embedology_Sauer_et_al_1991} show that almost every time-delay map is an embedding. Similarly,~\citet{Baranski_et_al_2020} show that the time-delay map formed by a polynomial perturbation of a given locally Lipschitz observation function is almost surely an embedding.~\citet{stark_et_al_1997, stark_et_al_2003} extend the~\citet{embedding_takens_1981} embedding theorem to forced
systems and stochastic systems.~\citet{Casdagli_et_al_1991} use statistical methods to account for the effect of iid observational noise on the state space reconstruction. Recently,~\citet{BotvinickGreenhouse_et_al_2025} established a measure-theoretic generalization using optimal transport theory. These results all trace back to the classical~\citet{whitney_embedding_1936} embedding theorem from differential topology. 

Related results also appear in parameter identification and observability.~\citet{aeyels_1981_2,aeyels_1981} proved that the observability of a $d$-dimensional dynamical system can be achieved by taking $2d+1$ exact measurements at random times. For analytically parametrized dynamical systems,~\citet{sontag_2003} proved that $2p+1$ generic experiments are sufficient to distinguish between distinct $p$-dimensional parameter values. However, these results, unlike ours, are restricted to dynamical systems \textit{without noise}.

The result from nonlinear dynamics most closely related to our work is the Fractal Whitney Embedding Prevalence Theorem from \citet{embedology_Sauer_et_al_1991}, so we discuss it in more detail. We begin by stating the five \textit{genericity axioms} introduced by \citet{prevalence_ott_yorke_2005}. For a topological vector space $V$, it is desirable that a definition of ``generic'' satisfies:
\begin{enumerate}
\item A generic subset of $V$ is dense in $V$.
\item If $F \supset E$ and $E$ is generic, then $F$ is generic.
\item A countable intersection of generic sets is generic.
\item Every translate of a generic set is generic.
\item A subset $E$ of finite-dimensional Euclidean space is generic if and only if $E$ is a set of full Lebesgue measure.
\end{enumerate}

The topological notion of genericity uses \textit{residual sets}, which are countable intersections of open, dense sets. By the Baire category theorem, residual subsets of Baire spaces are dense. However, the topological notion of genericity does not satisfy Axiom 5 because the Lebesgue measure of these residual sets can be zero. Thus, a measure-theoretic notion of genericity is required.

Unfortunately, the notion of ``probability one'' in finite-dimensional spaces does not extend in a canonical way to infinite-dimensional spaces. Therefore, since the space of all smooth maps is an infinite-dimensional space, it is difficult to assert that ``almost every'' smooth map has a given property; see~\citet{embedology_Sauer_et_al_1991} and~\citet{prevalence_Hunt_et_al_1992} for more discussion. To formalize the notion that ``almost every'' map has a given property, the idea of \textit{prevalence} is used.

\begin{definition}[Prevalence~\citep{prevalence_ott_yorke_2005}]\label{defn:prevalence} Let $V$ be a completely metrizable topological vector space. A Borel set $E\subset V$ is said to be \textit{prevalent} if there exists a Borel measure $\mu$ on $V$ such that: \begin{enumerate}
    \item $0<\mu(C)<\infty$ for some compact subset $C$ of $V$.
    \item The set $E+v$ has full $\mu$-measure (complement of $E+v$ has measure zero) for all $v\in V$.
\end{enumerate}
\end{definition}

If $E\subset V$ contains a prevalent Borel set, then it is common to say that ``almost every'' element of $V$ lies in $E$.

The embedding result from \citet{embedology_Sauer_et_al_1991} involves the notion of an \textit{immersion}.

\begin{definition}[Immersion~\citep{riemannian_geometry_Gallot_et_al}]\label{def:immersion} 
    A smooth map  
    $\Phi:\mathbb{R}^p\xrightarrow[]{}\mathbb{R}^k$ is an \textit{immersion} at $\theta\in\mathbb{R}^p$ if its differential map $D_{\theta}\Phi$ at $\theta$ is injective from $\mathbb{R}^p$ to $\mathbb{R}^k$.
\end{definition}

The following result establishes that ``almost every'' smooth map (in the sense of prevalence from Definition~\ref{defn:prevalence}) from $\mathbb{R}^p$ to $\mathbb{R}^k$, $k \geq 2p+1$, has the following properties.

%%% Note: if we wanted to, we could use random features that are dense in C^1 (e.g. linear combinations of RFFs, random neural networks, etc) and then conclude that embeddings are open and dense in this class of functions, because the set of embeddings in C^1 is open and dense by Whitney 1936.  also this would probably require a sieve-type argument, in which the complexity of the class of random functions that we sample from grows in complexity with n (i.e. more terms in the linear combination), because only the limiting function class is dense in C^1. Overall, this seems unnecessary because the current method works well without using linear combinations
%%%% More importantly, we are explicitly arguing that dense is not good enough, which is why we use prevalence for a probability 1 statement.  

\begin{theorem}[Fractal Whitney Embedding Prevalence Theorem~\citep{embedology_Sauer_et_al_1991}]\label{thm:fractal_whitney_embed_prev_thm} Let $\Theta$ be a compact subset of $\mathbb{R}^p$ of box-counting dimension $p'$, and let $k$ be an integer greater than $2p'$. For almost every smooth map $\Phi:\mathbb{R}^p\xrightarrow[]{}\mathbb{R}^k$:
\begin{enumerate}
    \item $\Phi$ is one-to-one on $\Theta$.
    \item $\Phi$ is an immersion on each compact subset of a smooth manifold contained in $\Theta$.
\end{enumerate}
\end{theorem}

%%% Note: the smaller class of functions does not (necessarily) have the embedding property almost surely
This result cannot be used for our purposes, though, because we cannot sample ``generic'' smooth maps without predetermining a class of functions. There are, however, function classes that are easy to sample from and still rich enough to characterize distributions. For instance, the expectations of Fourier basis functions are rich enough to uniquely identify the distributions of random variables taking values in Euclidean space.

Next, in Section~\ref{section:results}, we establish that the expectations of $2p+1$ generic random Fourier features suffice to identify models with a $p$-dimensional parameter. Section~\ref{subsection:id_stationary_dyn_mod} shows this for models of (asymptotically) stationary stochastic processes, such as deterministic dynamical systems observed through noise and models with process noise. Section~\ref{subsection:id_nonstationary_dyn_mod} presents an analogous result for nonstationary processes, such as ODEs observed through noise. 
\begin{remark}[Relationship to companion paper] The identification results in Section~\ref{section:results} were introduced in our companion paper \citep{random_features_estimation_paper}, which focuses on statistical estimation theory using the concept of physical dependence \cite{rosenblatt_1961,wiener_1966,wu_funct_dep_meas}. In contrast, this paper isolates the assumptions needed for identification and avoids statistical jargon. Also, this paper connects our identification results to previous works from nonlinear dynamics. We hope that, by making these identification results accessible to the physics and applied mathematics communities, this paper will help lead to further results in nonlinear dynamics.
\end{remark}

\begin{figure*}
\centering

\begin{tabular}{@{}cc@{}}
\textbf{(a)} & \textbf{(b)} \\[1mm]
\includegraphics[width=0.48\textwidth]{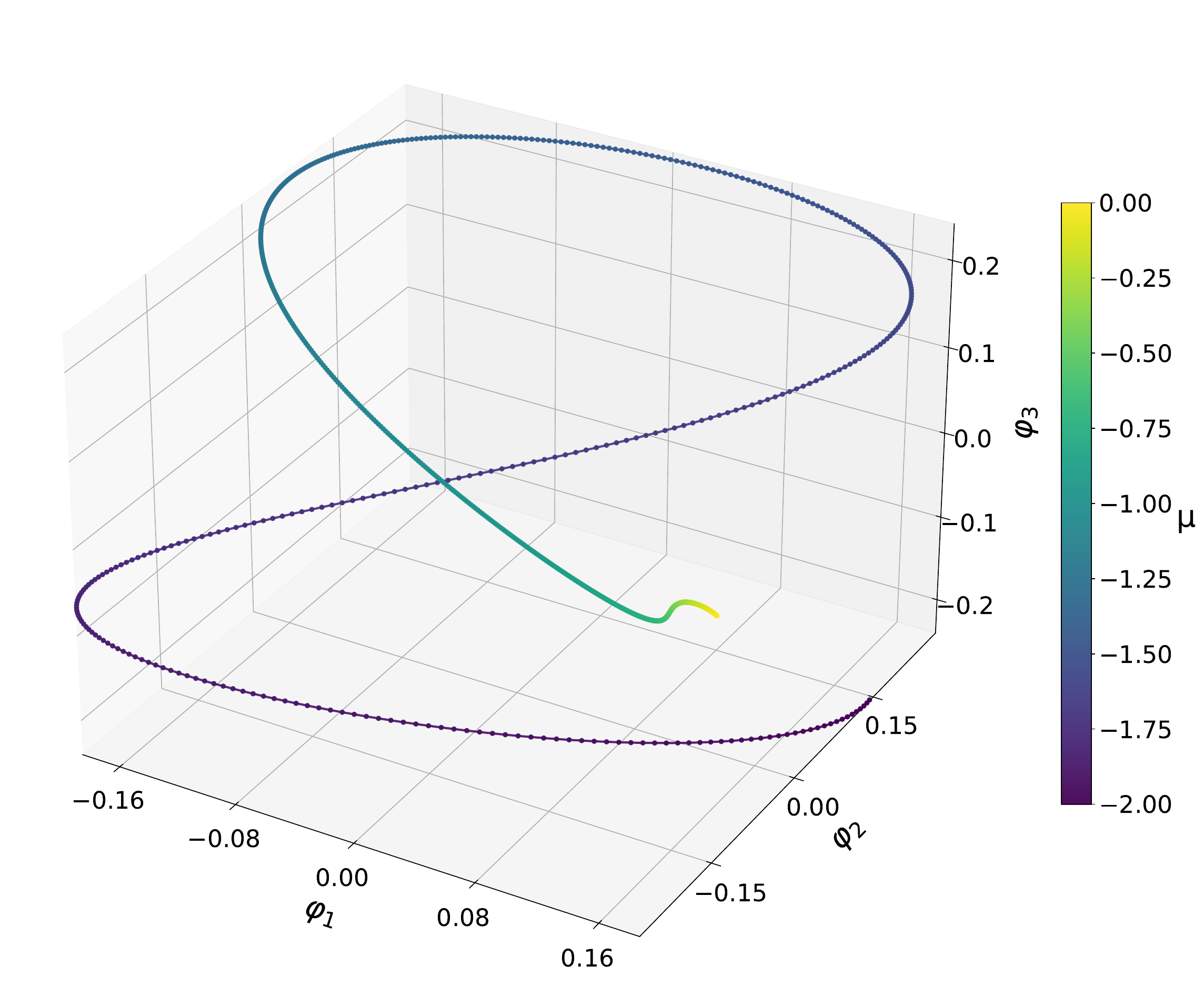}
&
\includegraphics[width=0.48\textwidth]{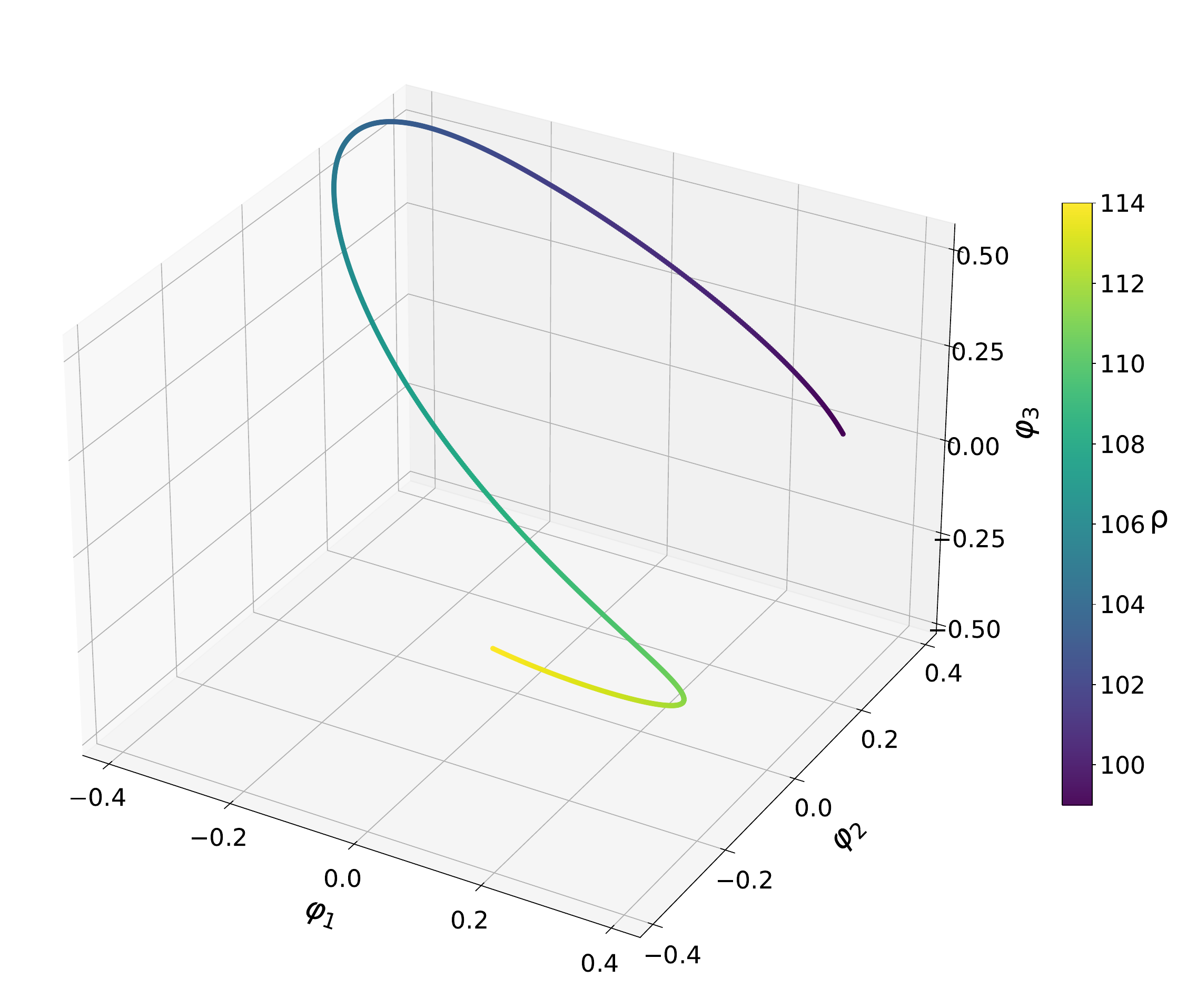}
\end{tabular}

\caption{The same $k=2p+1=3$ random features
$\varphi=(\varphi_1,\varphi_2,\varphi_3)$ as in
Eq.~\eqref{eqn:random_Fourier_features_order_m} identify every parameter
value in both the Gaussian distribution model and the Lorenz-63 model.\\
(a) Let $(X_t^{(r)}(\mu))_{r=1}^{s}$, $s=10$, be
$X_t^{(r)}(\mu)\overset{\mathrm{iid}}{\sim}
N((\mu,\mu^2,\mu^3)^{\top},I_3)$,
$r=1,\ldots,s$. The plot shows
$s^{-1}\sum_{r=1}^s \varphi(X_t^{(r)}(\mu))$ as a function of the mean
parameter $\mu$. Color denotes $\mu\in[-2,0]$.\\
(b) Let $(X_t^{(r)}(\rho))_{r=1}^{s}$, $s=10$, be the noisily observed
Lorenz-63 model as in Eq.~\eqref{eq:lorenz63_obs_full} at time $T=1$
with fixed parameter values $\sigma=10$, $\beta=8/3$, and
$\lambda^{(1)}=\lambda^{(2)}=\lambda^{(3)}=1$. The plot shows
$s^{-1}\sum_{r=1}^s \varphi(X_t^{(r)}(\rho))$ as a function of the
Rayleigh number $\rho$. Color denotes $\rho\in[99,114]$.}
\label{fig:estimation_procedure_and_iid_gaussian_feature_trajectory_3d}
\end{figure*}

% The result will not use the WBW framework, it will just be the identification principle in the most general form. In the cover letter, should mention the differences between these results assuming other paper is under review at another journal.

\section{Results}\label{section:results}

\subsection{Setting and notation}\label{subsection:setting_and_notation}

For a vector $x\in\mathbb{R}^d$, denote the $\ell^q$ norm by $\left\|x\right\|_q$ and the Euclidean norm by $\left\|x\right\| = \left\|x\right\|_2$. For a random vector $X$ with distribution depending on $\theta\in\Theta$, $\left\|X\right\|_{\mathcal{L}^q(\theta)}=\left(\mathbb{E}_{\theta}\left\|X\right\|^q\right)^{1/q}$, where $\mathbb{E}_{\theta}(\cdot)$ denotes expectation with respect to this $\theta$-dependent distribution. For any natural number $j\in\mathbb{N}$, denote $[j]=\{1,2,\ldots,j\}$. For a sequence of random vectors $X_t$, $t=1,\ldots,n$, each taking values in $\mathbb{R}^d$, we denote a subsequence as $X_{t_1:t_2}=(X_{t_1},\ldots,X_{t_2})^{\top}$, which takes values in $\mathbb{R}^{(t_2-t_1+1) \times d}$ for some $t_1,t_2 \in [n]$, $t_1\leq t_2$. For real numbers $x,y\in\mathbb{R}$, denote $x \land y = \min(x,y)$ and $x\lor y = \max(x,y)$.
% Note: \mathcal{L}^q(\theta) instead of L^q because using L for lag 

We observe $X_{t}^{\mathrm{obs}}$, $t=1,\ldots,n$, taking values in $\mathbb{R}^d$ for some dimension $d\in\mathbb{N}$ and for some sample size $n\in\mathbb{N}$. The observed sequence is assumed to come from a known parametric model with an unknown $p$-dimensional parameter vector $\theta_0\in\Theta\subset \mathbb{R}^{p}$, $p\in\mathbb{N}$. For any value of $\theta$, we denote the corresponding realizations of the time series by $X_{1:n}^{(r)}(\theta)$, $r=0,1,\ldots,s$ for some $s\in\mathbb{N}$, where we reserve the index $r=0$ for the observed time series, i.e., $X_{1:n}^{\mathrm{obs}}\equiv X_{1:n}^{(0)}(\theta_0)$. To refer to the $i$-th dimension of the $r$-th realization, we use the notation $X_{1:n}^{(r)(i)}$. When we do not need to refer to a particular realization or dimension, we drop the superscript.

\subsection{Random Fourier features}\label{subsection:random_Fourier_features}

We aim to estimate the $p$-dimensional parameter vector $\theta_0$ by matching the values of $2p+1$ random features of the simulated and observed data. We randomly draw $k=2p+1$ functions $\varphi_1,\ldots,\varphi_k$ from a distribution over a suitably nice function class $\mathcal{F}$, where each function $\varphi_i:\mathbb{R}^{(m+1) \times d}\xrightarrow[]{} \mathbb{R}$ takes a length $m+1$ subsequence of the $d$-dimensional time series as input. The number of lags $m$ used in the random features is chosen based on the number of lags $m^{\ast}$ in the transition function of the dynamic model.

Let $x_1,\ldots,x_{m+1}$ be vectors in $\mathbb{R}^d$ and denote $x=(x_1,\ldots,x_{m+1})^{\top}\in\mathbb{R}^{(m+1)\times d}$. Define 
\begin{equation}\label{eqn:random_Fourier_features_order_m}\varphi_{i}(x)=\cos\left(\sum_{j=1}^{m+1} \Omega_{i,j} \cdot x_{j} + \alpha_i\right),\end{equation} where $\Omega_{i,j} \overset{\mathrm{iid}}{\sim} N(0,I_d)$ and $\alpha_i \overset{\mathrm{iid}}{\sim} U(-\pi,\pi)$ for $i=1,\ldots,k$ and $j=1,\ldots,m+1$. Denote the parameters for the $i$-th random Fourier feature, $\varphi_i$, by \begin{equation}\label{eqn:parameters_for_the_random_features}\vartheta_i=(\Omega_{i,1}^{\top},\ldots,\Omega_{i,m+1}^{\top},\alpha_i)^{\top},\end{equation} and let $\vartheta=(\vartheta_1,\ldots,\vartheta_k)$, which is a $\mathbb{R}^{(d(m+1)+1)\times k}$-valued random variable. Denote \begin{equation}\label{eqn:all_k_random_Fourier_features_order_m}\varphi=(\varphi_1,\ldots,\varphi_k),\end{equation} so that $\varphi:\mathbb{R}^{(m+1) \times d}\xrightarrow[]{} \mathbb{R}^k$.

\subsection{Identification of stationary dynamic models}\label{subsection:id_stationary_dyn_mod}

The identification result presented here is for stochastic process models with meaningful stationary distributions. Denote by $P_{\theta,t}^{(r)}$ the distribution of $\left[X_{t-m}^{(r)}(\theta),\ldots,X_t^{(r)}(\theta)\right]^{\top}$. We require the average of the distributions $(P_{\theta,t}^{(r)})_{t\in\mathbb{N}}$ over time to converge weakly to a limiting distribution $P_{\theta}$, \textit{uniformly} over the parameter values $\theta\in\Theta$. This condition allows us to accommodate models with unknown, possibly random, initial values.
% Note: for example, the logistic / Henon map models

\begin{assumption}\label{asmpt:weak_convergence_of_time_avg_of_distributions_discrete_time} 
    There exists a family of limiting distributions $(P_{\theta})_{\theta\in\Theta}$ such that, for all $r=0,1,\ldots,s$, we have, as $n\xrightarrow[]{}\infty$,
    $$\underset{\theta\in\Theta}{\sup} \ \left|
\begin{aligned}
 \int_{\mathbb{R}^{(m+1)\times d}}\phi(x) \, d\bar{P}_{\theta,n}^{(r)}(x) - \int_{\mathbb{R}^{(m+1)\times d}}\phi(x) \, dP_{\theta}(x)\end{aligned} \right|\xrightarrow[]{} 0,$$ for all continuous, bounded functions $\phi:\mathbb{R}^{(m+1)\times d}\xrightarrow[]{}\mathbb{R}$, where $\bar{P}_{\theta,n}^{(r)}=\frac{1}{n-m}\sum_{t=m+1}^n P_{\theta,t}^{(r)}$.
    
\end{assumption}
% Note: this will also imply weak convergence when "late starting" i.e. starting at t=t_0+1 rather than t=1

We introduce mappings $\Phi:\Theta\xrightarrow[]{}\mathbb{R}^k$, given by \begin{equation}\label{eqn:limiting_expectation_random_features_discrete_time} \Phi(\theta)=\int_{\mathbb{R}^{(m+1)\times d}}\varphi(x) \, dP_{\theta}(x), \quad \theta \in \Theta.\end{equation}

Next, we require some constraints on the geometric complexity of the parameter space $\Theta$. For the following assumption, we need the definition of a semialgebraic set. We simplify the definition to Euclidean spaces.

\begin{definition}[Definition 2.1.4 in \citep{bochnak_coste_roy_real_algebraic_geometry}]\label{def:semialgebraic_set}
Let $p,a,b_1,\ldots,b_a \in \mathbb{N}$. A \textit{semialgebraic subset} of $\mathbb{R}^p$
is a subset of the form
\[
\bigcup_{i=1}^{a}\bigcap_{j=1}^{b_i}
\left\{x \in \mathbb{R}^p
\mid f_{i,j}(x) \mathbin{*_{i,j}} 0\right\},
\]
where $f_{i,j} \in \mathbb{R}[x_1,\ldots,x_p]$, $*_{i,j}$ is either
$<$ or $=$, for $i=1,\ldots,a$ and $j=1,\ldots,b_i$, and $\mathbb{R}[x_1,\ldots,x_p]$ is the ring of polynomials in
the variables $x_1,\ldots,x_p$ with coefficients in $\mathbb{R}$.
\end{definition}

%We present some additional notation for the next assumption. Denote by $\mathcal{P}\left(\mathbb{R}^{(m+1)\times d}\right)$ the space of probability measures on $\mathbb{R}^{(m+1)\times d}$, and 

Denote by $\psi_{\theta}(\omega)$ the characteristic function of $P_{\theta}$ evaluated at $\omega\in\mathbb{R}^{(m+1)\times d}$.

\begin{assumption}\label{asmpt:statistical_model_discrete_time}
The following conditions hold:
\begin{enumerate} 
    \item $\Theta$ is a compact, semialgebraic subset of $\mathbb{R}^{p}$ with non-empty interior. 
    \item The map $\theta\mapsto P_{\theta}$ is injective.  
    \item The real and imaginary parts of the map $(\theta,\omega)\mapsto \psi_{\theta}(\omega)$ each admit a jointly real analytic extension to $O \times \mathbb{R}^{(m+1)\times d}$, where $O$ is open and $\Theta \subset O\subset \mathbb{R}^p$. % i.e., $O$ $is an open neighborhood of $\Theta$ 
\end{enumerate}    
\end{assumption}

For example, for some $a_j,b_j\in\mathbb{R}$ with $a_j < b_j$, $j=1,\ldots,p$, we may take   $$\Theta=\prod_{j=1}^p [a_j,b_j].$$

The following result establishes that $2p+1$ generic random Fourier features are sufficient to identify a dynamic model with a $p$-dimensional parameter.

\begin{theorem}\label{thm:injective_discrete_time}
Suppose Assumptions~\ref{asmpt:weak_convergence_of_time_avg_of_distributions_discrete_time}-\ref{asmpt:statistical_model_discrete_time} hold. Then, with probability one over the draw of random parameters $(\vartheta_1,\ldots,\vartheta_k)$ for the $k=2p+1$ random Fourier features $(\varphi_1,\ldots,\varphi_k)$, we have that the mapping $\theta\mapsto\Phi(\theta)$ is one-to-one, where $\Phi(\theta)$, $\theta\in\Theta$, are the limiting expectations of the random Fourier features.
\end{theorem}

Suppose the randomly chosen functions from Eq.~\eqref{eqn:all_k_random_Fourier_features_order_m} yield a function from Eq.~\eqref{eqn:limiting_expectation_random_features_discrete_time} that is one-to-one. Indeed, by Theorem~\ref{thm:injective_discrete_time}, they will, almost surely. Then we can distinguish parameters $\theta_1 \neq\theta_2$ by looking at the expectation values of the random features, $\Phi(\theta_1)$ and $\Phi(\theta_2)$. In particular, the discrepancy function \begin{equation}\label{eqn:pop_discrep_fn_TA} Q^{\mathrm{TA}}(\theta) =\Phi(\theta_0) - \Phi(\theta),\end{equation} will be nonzero for $\theta\neq\theta_0$, which establishes that the parameters are identifiable. 
%%% Note: In our companion paper \citep{random_features_estimation_paper}, under additional assumptions, we use this result to establish the consistency of an estimator of $\theta_0$ based on averaging random Fourier features over time.

\subsection{Identification of nonstationary dynamic models}\label{subsection:id_nonstationary_dyn_mod}

For many nonstationary processes, e.g., ODEs with observational noise, the time-average of the random features may not converge to a limiting value that uniquely identifies the parameters. Hence, a \textit{local} approach is required. In this setting, as the sample size $n$ grows, we gain an increasing number of observations related to each local structure of the nonstationary process. That is, \begin{equation}\label{eqn:nonstationary_process_rescaled_time} X_t^{\mathrm{obs}}=\widetilde{X}_{t}^{\mathrm{obs}}(t/n),\quad X_t^{(r)}(\theta)=\widetilde{X}_{t}^{(r)}(t/n,\theta),\end{equation} for some random vectors $\widetilde{X}_{t}^{\mathrm{obs}}(u)$, $\widetilde{X}_{t}^{(r)}(u,\theta)$ that are well-defined at every \textit{rescaled} time $u\in [0,1]$. At every fixed $u\in [0,1]$, each sequence $(\widetilde{X}_{t}^{\mathrm{obs}}(u))_{t\in\mathbb{Z}}$ and $(\widetilde{X}_{t}^{(r)}(u,\theta))_{t\in\mathbb{Z}}$ is a stationary process. Nonstationarity in the processes from Eq.~\eqref{eqn:nonstationary_process_rescaled_time} occurs due to changes in the data-generating mechanism over rescaled time $u$. Note that $X_t^{\mathrm{obs}}$ and $X_t^{(r)}(\theta)$ both depend on $n$, although we suppress this to simplify the notation.

%%% Note: Since we are not using the the causal representation G_u from the estimation paper, we need to define the local distribution as the local approximation.

For each $\theta\in\Theta$ and $u\in [0,1]$, let $P_{\theta,u}$ be the \textit{locally stationary approximation} of the distribution of the nonstationary process at rescaled time $u$. That is, let $P_{\theta,u}$ be the distribution of $\left[\widetilde{X}_{t-m}(u,\theta),\ldots,\widetilde{X}_{t}(u,\theta)\right]^{\top}$ and, for any sequence of times $(t_n)_{n\in\mathbb{N}}$ with $t_n\in\{m+1,\ldots,n\}$ such that $t_n/n \xrightarrow[]{} u$, as $n\xrightarrow[]{}\infty$, the distribution of $\left[X_{t_n-m}(\theta),\ldots,X_{t_n}(\theta)\right]^{\top}$ converges weakly to $P_{\theta,u}$. Note that $P_{\theta,u}$ does not depend on the simulation index $r$ or the time $t$. For the technical details about these \textit{locally stationary processes}, see the companion paper~\citep{random_features_estimation_paper}.

We introduce mappings $\Phi_u:\Theta\xrightarrow[]{}\mathbb{R}^k$, $u\in [0,1]$, given by \begin{equation}\label{eqn:original_random_feature_expectation_cont_time} \Phi_u(\theta)= \int_{\mathbb{R}^{(m+1)\times d}}\varphi(x) \, dP_{\theta,u}(x).\end{equation} Note that, in general, we may have $P_{\theta_1,u}=P_{\theta_2,u}$ for $\theta_1\neq\theta_2$, so it is often impossible to injectively map $\Theta$ into $\mathbb{R}^k$ using $\Phi_u$ at a particular rescaled time $u\in [0,1]$.

For the next assumption, the definition of a semialgebraic mapping is required. We simplify the definition from \citet{bochnak_coste_roy_real_algebraic_geometry} to Euclidean spaces.

\begin{definition}[Definition 2.2.5 in \citep{bochnak_coste_roy_real_algebraic_geometry}]\label{def:semialgebraic_map}
Let $p_1,p_2 \in \mathbb{N}$, and let
$A \subset \mathbb{R}^{p_1}$ and $B \subset \mathbb{R}^{p_2}$ be two
semialgebraic sets (Definition~\ref{def:semialgebraic_set}). A mapping $g \colon A \to B$ is semialgebraic if
its graph $G(g)=\{(x,g(x)):x\in A\}$ is semialgebraic in $\mathbb{R}^{p_1+p_2}$.
\end{definition}

The following reparametrization assumption is needed for identifiability.

\begin{assumption}\label{asmpt:reparametrization_continuous_time} For each $u\in [0,1]$, there exists a Lipschitz, semialgebraic mapping $g_u:\Theta\xrightarrow[]{}\mathbb{R}^{\tilde{p}}$ defined by $g_u(\theta)=\tilde{\theta}_u$, where $\tilde{\Theta}_u=\{g_u(\theta):\theta\in\Theta\}\subset\mathbb{R}^{\tilde{p}}$, $\tilde{p}\in \mathbb{N}$, and there exists a reparametrization such that $$\widetilde{X}'_{t}(u,\tilde{\theta}_u)=\widetilde{X}_{t}(u,\theta).$$ Furthermore, assume that the map $(u,\theta)\mapsto g_u(\theta)$ is jointly Borel measurable.

\end{assumption}

We may substantially weaken the requirement that $g_u$ is semialgebraic (Definition~\ref{def:semialgebraic_map}). In fact, we only require $g_u$ to be $\sigma$-subanalytic; see Appendix~\ref{section:proofs} for the technical details.

Let $\tilde{P}_{\tilde{\theta}_u,u}$ be the distribution of $\left[\widetilde{X}'_{t-m}(u,\tilde{\theta}_u),\ldots,\widetilde{X}'_{t}(u,\tilde{\theta}_u)\right]^{\top}$. We introduce mappings $\tilde{\Phi}_u:\tilde{\Theta}_u\xrightarrow[]{}\mathbb{R}^k$ defined as \begin{equation}\label{eqn:reparam_random_feature_expectation}\tilde{\Phi}_u(\tilde{\theta}_u) = \int_{\mathbb{R}^{(m+1)\times d}}\varphi(x) \, d\tilde{P}_{\tilde{\theta}_u,u}(x),\end{equation} 
which are the \textit{local} expectations of the random Fourier features at rescaled time $u\in [0,1]$.

% Denote by  $\mathcal{P}\left(\mathbb{R}^{(m+1)\times d}\right)$ the space of probability measures on $\mathbb{R}^{(m+1)\times d}$, and 
Denote by $\tilde{\psi}_{\tilde{\theta}_u,u}(\omega)$ the characteristic function of $\tilde{P}_{\tilde{\theta}_u,u}$ evaluated at $\omega\in\mathbb{R}^{(m+1)\times d}$.

\begin{assumption}\label{asmpt:statistical_model_continuous_time}
The following conditions hold for all rescaled times $u\in [0,1]$: 
\begin{enumerate}
    \item $\tilde{\Theta}_u\subset \mathbb{R}^{\tilde{p}}$ is compact. $\Theta\subset \mathbb{R}^{p}$ is compact and semialgebraic with non-empty interior.
    \item The map $\tilde{\theta}_u\mapsto \tilde{P}_{\tilde{\theta}_u,u}$ is injective.
    \item The real and imaginary parts of the map $(\tilde{\theta}_u,\omega)\mapsto\tilde{\psi}_{\tilde{\theta}_u,u}(\omega)$ each admit a jointly real analytic extension to $\tilde{O}_u \times \mathbb{R}^{(m+1)\times d}$, where $\tilde{O}_u$ is open and $\tilde{\Theta}_u\subset \tilde{O}_u \subset\mathbb{R}^{\tilde{p}}$.
% i.e., $\tilde{O}_u$ $is an open neighborhood of $\tilde{\Theta}_u$ 
\end{enumerate}
\end{assumption}

The next assumption is used to show that the mapping $\tilde{\theta}_u\mapsto\tilde{\Phi}_u(\tilde{\theta}_u)$, $u\in [0,1]$, is one-to-one for $2p+1$ generic random Fourier features at almost every rescaled time $u$.

\begin{assumption}\label{asmpt:measurability} The following conditions hold:
\begin{enumerate}
    \item The set-valued map $u\mapsto \tilde{\Theta}_u$ is weakly Borel measurable. 
    \item There exists a jointly Borel measurable function $\bar{\psi}:[0,1] \times \mathbb{R}^{\tilde{p}} \times \mathbb{R}^{(m+1)\times d}\xrightarrow[]{}\mathbb{C}$ such that $\bar{\psi}(u,\tilde{\theta}_u,\omega)= \tilde{\psi}_{\tilde{\theta}_u,u}(\omega)$ for all $u\in [0,1]$, $\tilde{\theta}_u\in\tilde{\Theta}_u$, and $\omega\in\mathbb{R}^{(m+1)\times d}$, and is continuous in $\tilde{\theta}_u$ for each fixed $(u,\omega)$.     
\end{enumerate}
    
\end{assumption}

The next result is critical for establishing the identifiability of the parameters.

\begin{theorem}\label{thm:injective_continuous_time} Suppose Assumptions~\ref{asmpt:reparametrization_continuous_time}-\ref{asmpt:measurability} hold. Then, with probability one over the draw of random parameters $(\vartheta_1,\ldots,\vartheta_k)$ for the $k=2p+1$ random Fourier features $(\varphi_1,\ldots,\varphi_k)$, we have that the mappings $\tilde{\theta}_u \mapsto \tilde{\Phi}_u(\tilde{\theta}_u)$, $u\in [0,1]$, are one-to-one for almost every $u\in [0,1]$, where $\tilde{\Phi}_u(\tilde{\theta}_u)$, $\tilde{\theta}_u\in\tilde{\Theta}_u$, are the limiting local expectations of the random Fourier features.
\end{theorem}

Lastly, the following assumption is required to distinguish parameters based on the discrepancy function \begin{equation}\label{eqn:pop_discrep_RW} Q^{\mathrm{RW}}(\theta) =\int_0^1 \norm{\Phi_u(\theta_0) - \Phi_u(\theta)}^2 du.\end{equation}

\begin{assumption}\label{asmpt:distinguish_paths_of_distributions_continuous_time} For all $\theta_1,\theta_2\in\Theta$, if $\theta_1\neq\theta_2$, then there exists a subset $\mathcal{U}_{\theta_1,\theta_2}\subseteq [0,1]$ with positive Lebesgue measure such that $P_{\theta_1,u}\neq P_{\theta_2,u}$ for all rescaled times $u\in \mathcal{U}_{\theta_1,\theta_2}$.
\end{assumption}

If the expectations of the $2p+1$ random Fourier features are one-to-one at almost all times (by Theorem~\ref{thm:injective_continuous_time}, they will be, almost surely), then we can distinguish parameters $\theta_1 \neq\theta_2$ by comparing the local expectations of the random features $\Phi_u(\theta_1)$, $\Phi_u(\theta_2)$ at almost any rescaled time in $\mathcal{U}_{\theta_1,\theta_2}$. Moreover, $Q^{\mathrm{RW}}(\theta)$ from Eq.~\eqref{eqn:pop_discrep_RW} will be nonzero for $\theta\neq\theta_0$, which establishes the identifiability of the parameters. 
%%% Note: In our companion paper \citep{random_features_estimation_paper}, under additional regularity conditions, we use the identification result from Theorem~\ref{thm:injective_continuous_time} to establish the consistency of an estimator of $\theta_0$ based on rolling-window averages. 

\begin{remark}[Restrictiveness of assumptions]\label{rmk:restrictiveness_of_asmpts} The assumptions we use to establish Theorems~\ref{thm:injective_discrete_time} and~\ref{thm:injective_continuous_time} are more restrictive than necessary. We leave the problem of finding necessary and sufficient conditions for identification via random Fourier features for future work.
\end{remark}

\section{Applications to atmospheric convection models}\label{section:simulations}

We demonstrate the practical relevance of our identification results by estimating (or calibrating) two prototypical models of atmospheric convection; namely, the H\'{e}non map and Lorenz-63 system. For the following discussion, denote the $i$-th random feature at time $t=m+1,\ldots,n$ by  $$f_{t,i}^{\mathrm{obs}}=\varphi_i\left(X_{t-m:t}^{\mathrm{obs}}\right), \quad f_{t,i}^{(r)}(\theta)=\varphi_i\left( X_{t-m:t}^{(r)}(\theta)\right),$$ and write all random features at time $t$ as \begin{equation}\begin{aligned}\label{eqn:random_features_at_time_t} f_{t}^{\mathrm{obs}} & =\left(f_{t,1}^{\mathrm{obs}},\ldots,f_{t,k}^{\mathrm{obs}}\right), \quad  f_{t}^{(r)}(\theta) & =\left(f_{t,1}^{(r)}(\theta),\ldots,f_{t,k}^{(r)}(\theta)\right).\end{aligned}\end{equation}

\begin{remark}[Assumptions for Lorenz-63 and H\'{e}non map]
    Our theoretical framework can accommodate the Lorenz-63 system if $g_u$ is taken to be $\sigma$-subanalytic (see Appendix A) rather than semialgebraic (see Definition~\ref{def:semialgebraic_map}); see the discussion below Assumption~\ref{asmpt:reparametrization_continuous_time}. We include the H\'{e}non map as a numerical illustration because of its relationship to the Lorenz-63 system. However, as far as we know, it is not obvious which restrictions on the parameter space $\Theta$ are needed to ensure that the H\'{e}non map model satisfies Assumptions~\ref{asmpt:weak_convergence_of_time_avg_of_distributions_discrete_time} and~\ref{asmpt:statistical_model_discrete_time}.
\end{remark}

\subsection{H\'{e}non map model}\label{subsection:henon_map}

\citet{henon_1976} introduced a discrete-time model that exhibits chaotic behavior, which is a simplified model for the Poincar\'{e} section of the Lorenz-63 system~\citep{lorenz_1963}. For times $t=1,\ldots,n$, the dynamical system is defined by \begin{equation}\label{eq:henon_full}
\begin{aligned}
Z_{t}^{(1)} &= 1- a (Z_{t-1}^{(1)})^2 + Z_{t-1}^{(2)}, \\
Z_{t}^{(2)} &= b Z_{t-1}^{(1)},
\end{aligned}
\end{equation} where $Z_t^{(1)}$, $Z_t^{(2)}$ are the state variables. The unknown initial values $Z_0^{(1)}$ and $Z_0^{(2)}$ are sampled from $U[-0.7,0.7]$ and $U[-0.2,0.2]$ distributions, respectively. The components of the parameter vector are: $a$, which relates to the forcing or stress, and $b$, which relates to the damping or dissipation.

For times $t=1,\ldots,n$, we observe a noisy version of the first coordinate
\begin{equation}\label{eq:henon_obs_full}
\begin{aligned}
X_t &= Z_t^{(1)} + \sigma\varepsilon_t,
\end{aligned}
\end{equation} where $\varepsilon_t\overset{\mathrm{iid}}{\sim}N(0,1)$ and $\sigma$ controls the noise level. We use iid Gaussian noise to simplify the exposition, but in general the noise can be non-iid and non-Gaussian. We aim to estimate
\[
\begin{aligned}
a_0 &= 1.4, & b_0 &= 0.3, & 
\sigma_0 &= 0.1,
\end{aligned}
\]
using the following \textit{time-average} estimator, which was shown to converge in probability to the true parameter in our companion paper \citep{random_features_estimation_paper}.

Denote the time-averages of the random Fourier features by 
\begin{equation}\begin{aligned}\label{eqn:time_average_random_features}  F^{\mathrm{obs}} &= \frac{1}{n-m}\sum_{t=m+1}^n f_{t}^{\mathrm{obs}}, \quad  \bar{F}^{\mathrm{sim}}(\theta)&= \frac{1}{n-m}\sum_{t=m+1}^n \bar{f}_{t}^{\mathrm{sim}}(\theta), \end{aligned}\end{equation}  where $\bar{f}_{t}^{\mathrm{sim}}(\theta)=\frac{1}{s}\sum_{r=1}^s f_{t}^{(r)}(\theta)$ and $f_{t}^{\mathrm{obs}}$, $f_{t}^{(r)}(\theta)$ are from Eq.~\eqref{eqn:random_features_at_time_t}. The empirical discrepancy function corresponding to Eq.~\eqref{eqn:pop_discrep_fn_TA} is given by \begin{equation}\label{eqn:time_average_sample_discrepancy}\hat{Q}_n^{\mathrm{TA}}(\theta) =F^{\mathrm{obs}} - \bar{F}^{\mathrm{sim}}(\theta).\end{equation} The estimator $\hat{\theta}^{\mathrm{TA}}$ is obtained by minimizing $\left\|\hat{Q}_n^{\mathrm{TA}}(\cdot)\right\|$ with an optimization procedure. Specifically, we used the \texttt{differential\_evolution} solver from the \texttt{scipy.optimize} module in the SciPy Python library with $s=10$ simulations per parameter value.

\begin{figure}[h!]
\centering
\includegraphics[width=0.75\linewidth]{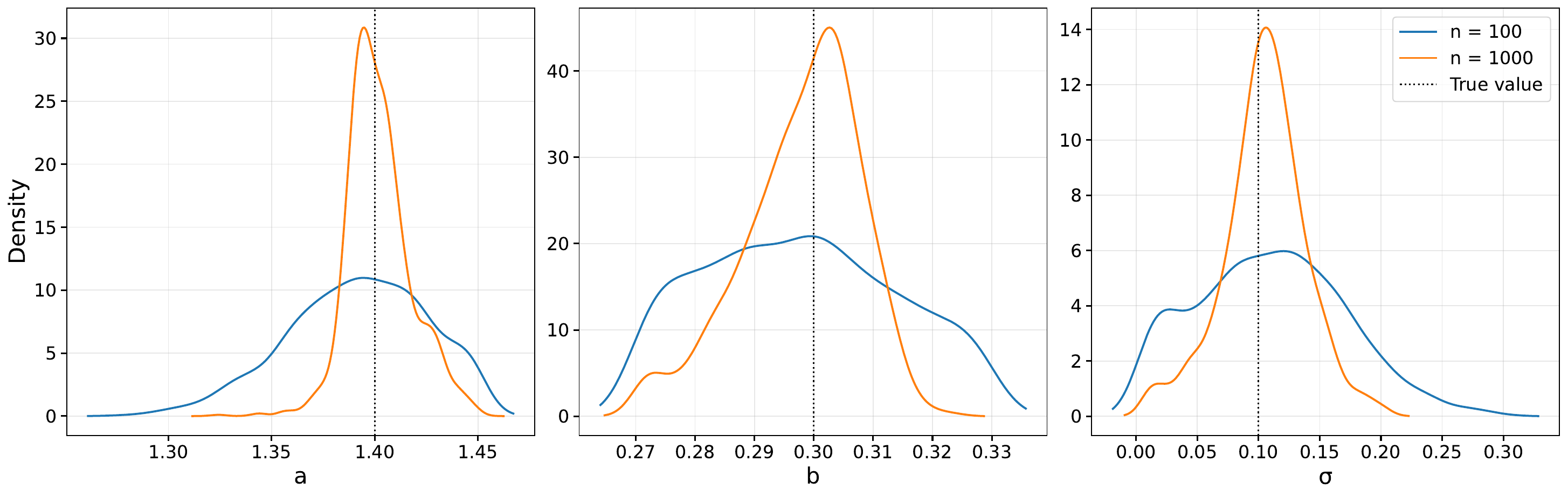}
\caption{\textbf{H\'{e}non map model.} Density of $1{,}000$ parameter estimates using $2p+1=7$ features.}
\label{fig:henon_estimate_densities_params0-5}
\end{figure}

\subsection{Lorenz-63 model}\label{subsection:lorenz63}

\citet{lorenz_1963} introduced a model for atmospheric convection that has become a canonical example of chaotic dynamics. For time horizon $T=1$, let $(Z^{(1)}(v),Z^{(2)}(v),Z^{(3)}(v))_{v\in[0,T]}$ be defined by the system of ODEs
\begin{equation}\label{eq:lorenz63_continuous}
\begin{aligned}
\frac{d Z^{(1)}}{dv}(v) &= \sigma(Z^{(2)}(v)-Z^{(1)}(v)), \\
\frac{d Z^{(2)}}{dv}(v) &= Z^{(1)}(v)(\rho-Z^{(3)}(v))-Z^{(2)}(v), \\
\frac{d Z^{(3)}}{dv}(v) &= Z^{(1)}(v)Z^{(2)}(v)-\beta Z^{(3)}(v),
\end{aligned}
\end{equation}
where $Z^{(1)}(v)$, $Z^{(2)}(v)$, $Z^{(3)}(v)$ are the state variables at time $v\in[0,T]$. $Z^{(1)}(v)$ is proportional to the intensity of the convection, $Z^{(2)}(v)$ relates to the horizontal temperature variation, and $Z^{(3)}(v)$ relates to the vertical temperature variation. The known initial values are $Z^{(1)}(0)=1$, $Z^{(2)}(0)=1$, $Z^{(3)}(0)=1$. The components of the parameter vector are: $\sigma$, which relates to the Prandtl number, $\rho$, which relates to the Rayleigh number, and $\beta$, which relates to the fluid layer's physical dimensions.

We assume equally spaced observations, but this is not necessary. For $t=1,\ldots,n$, we observe
\begin{equation}\label{eq:lorenz63_obs_full}
\begin{aligned}
X_t^{(1)} &= Z^{(1)}(Tt/n) + \lambda^{(1)}\varepsilon_t^{(1)}, \\
X_t^{(2)} &= Z^{(2)}(Tt/n) + \lambda^{(2)}\varepsilon_t^{(2)}, \\
X_t^{(3)} &= Z^{(3)}(Tt/n) + \lambda^{(3)}\varepsilon_t^{(3)},
\end{aligned}
\end{equation}
where $\varepsilon_t^{(1)},\varepsilon_t^{(2)},\varepsilon_t^{(3)}\overset{\mathrm{iid}}{\sim}N(0,1)$ and $\lambda^{(1)}$, $\lambda^{(2)}$, $\lambda^{(3)}$ control the noise level. We aim to estimate
\[
\begin{aligned}
\sigma_0 &= 10, & \rho_0 &= 100.5, & \beta_0 &= 8/3, \\
\lambda_0^{(1)} &= 1, & \lambda_0^{(2)} &= 1, & \lambda_0^{(3)} &= 1,
\end{aligned}
\] 
using the following \textit{rolling-window} estimator, which was shown to converge in probability to the true parameter in our companion paper \citep{random_features_estimation_paper}.

Denote the rolling averages of the random Fourier features up to time $t$ by \begin{equation}\begin{aligned}\label{eqn:rolling_window_random_features}  F_{t}^{\mathrm{obs}}&=\frac{1}{N_t}\sum_{j=(t-w)\lor (m+1)}^t f_{j}^{\mathrm{obs}},\quad  \bar{F}_{t}^{\mathrm{sim}}(\theta) &=  \frac{1}{N_t}\sum_{j=(t-w)\lor (m+1)}^t \bar{f}_{j}^{\mathrm{sim}}(\theta),
\end{aligned}\end{equation} where $N_t=N_t(w,m)=(w+1)\land (t-m)$ for some window size parameter $w\in\mathbb{N}$ and $\bar{f}_{t}^{\mathrm{sim}}(\theta)=\frac{1}{s}\sum_{r=1}^s f_{t}^{(r)}(\theta)$ and $f_{t}^{\mathrm{obs}}$, $f_{t}^{(r)}(\theta)$ are from Eq.~\eqref{eqn:random_features_at_time_t}. The empirical discrepancy function corresponding to Eq.~\eqref{eqn:pop_discrep_RW} is given by \begin{equation}\label{eqn:rolling_window_sample_discrepancy} \begin{aligned} \hat{Q}_n^{\mathrm{RW}}(\theta)&= \frac{1}{n-m}\sum_{t=m+\tau+L}^{n} \left[\left\|F_{t-L}^{\mathrm{obs}}-\bar{F}_{t-L}^{\mathrm{sim}}(\theta)\right\|^2 + K_{t}(\theta)\right],\\   K_{t}(\theta)&=2\left(F_{t-L}^{\mathrm{obs}}-\bar{F}_{t-L}^{\mathrm{sim}}(\theta) \right)^{\top}\left(\left[f_t^{\mathrm{obs}}-\bar{f}_{t}^{\mathrm{sim}}(\theta)\right]-\left[F_{t-L}^{\mathrm{obs}}-\bar{F}_{t-L}^{\mathrm{sim}}(\theta)\right]\right),\end{aligned}\end{equation} and where $\tau\in\mathbb{N}$ is an initial time-offset and $L\in\mathbb{N}$ is a lag. In our companion paper \citep{random_features_estimation_paper}, we establish theoretical guarantees for the rolling-window estimator by taking $w=w_n$, $\tau=\tau_n$, and $L=L_n$ to grow with the sample size $n$ at a particular rate. Specifically, the lag $L_n$ is set to $\lceil \frac{1}{10}\log(n)^2\rceil$, the window size $w_n$ is selected as the integer within $[n^{\frac{1}{2}},n^{\frac{3}{4}}]$ that minimizes the sum of squared distances $\sum_{t=m+1+L_n}^n \left\|F_{t-L_n}^{\mathrm{obs}}-f_t^{\mathrm{obs}}\right\|^2$, and the offset is set to $\tau_n=w_n$. We use the same settings for the experiments with the Lorenz-63 model here.

The estimator $\hat{\theta}^{\mathrm{RW}}$ is obtained by minimizing $\left|\hat{Q}_n^{\mathrm{RW}}(\cdot)\right|$ via optimization. For these experiments, we used the \texttt{differential\_evolution} solver from the \texttt{scipy.optimize} module with $s=10$ simulations per parameter value, and we used the Runge--Kutta 5(4) solver \texttt{RK45} from the \texttt{scipy.integrate} module, both from the SciPy Python library.

\begin{figure}[h!]
\centering
\includegraphics[width=1\linewidth]{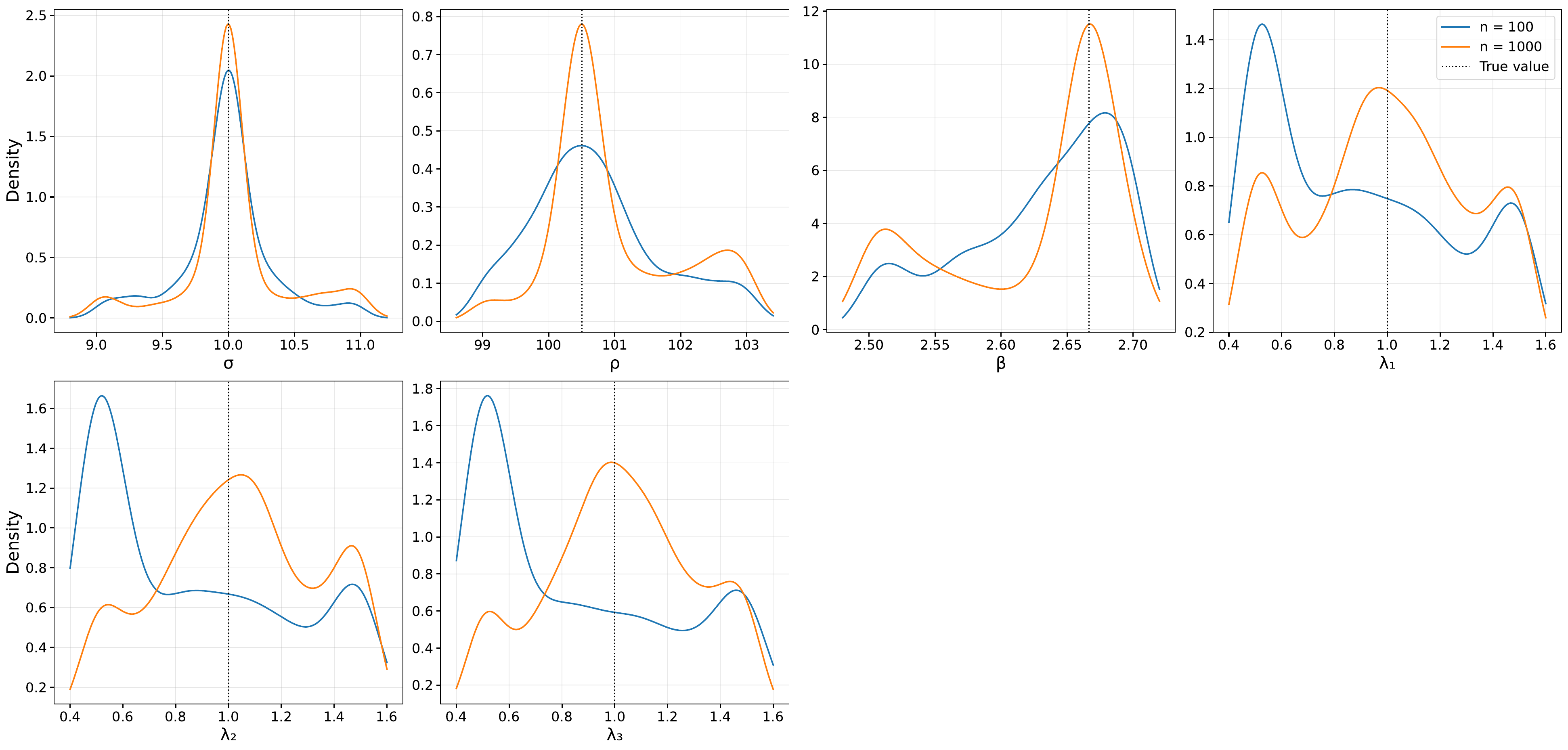}
\caption{\textbf{Lorenz-63 model.} Density of $1{,}000$ parameter estimates using $2p+1=13$ features.}
\label{fig:lorenz63_estimate_densities_params0-5}
\end{figure}

\section{Discussion}

We present identification results for dynamic models \textit{with noise}, which are inspired by previous results in nonlinear dynamics. It should be possible to extend our identification principle beyond dynamic models for time series. For example, it would be of interest to extend our ideas to parametric models for spatiotemporal data and network data. In general, the essential component is a class of random features for the dependence structure at hand.

Our identification results lead to broadly applicable estimation strategies based on matching $2p+1$ random features of the simulated and observed data. Existing simulation-based estimation approaches are costly, because they either require user-chosen summary statistics or computationally intensive neural network training for learning summary statistics. In contrast, our approach requires very little input from the user. To get standard errors, a model-based bootstrap procedure can be used. In future work, we will establish the asymptotic normality of both estimators and explain how to obtain valid confidence sets.

\appendix

\section{Proofs}\label{section:proofs}

\subsection{Preliminaries}\label{subsection:preliminaries}

In this subsection, we state several definitions and results from other papers. We use the notation originally used in those papers.

\begin{figure}[h!]
%\captionsetup{width=\linewidth, margin=0pt}
  \centering \includegraphics[width=1\linewidth]{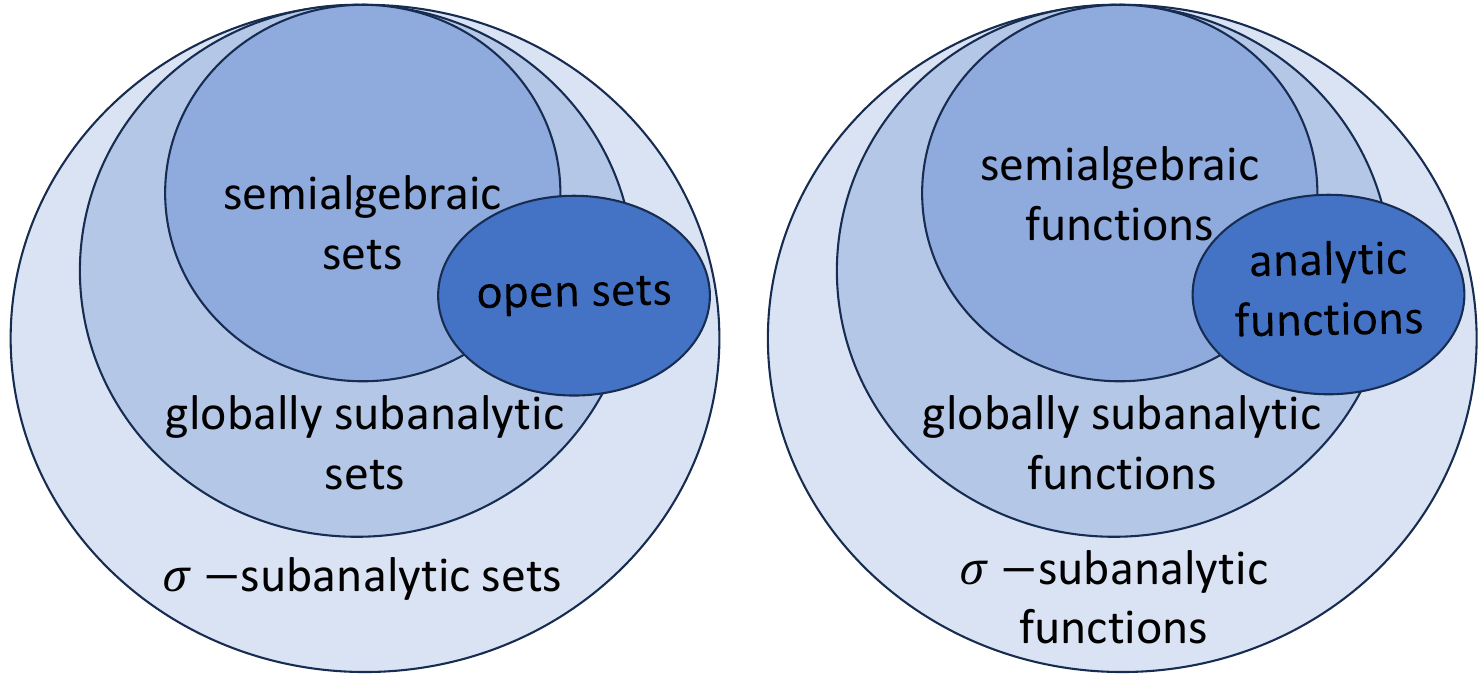}
  \caption{Classes of sets and functions (Figure 3 in \citet{amir_et_al_finite_witness_thm}).}\label{fig:diagram_of_subanalytic_sets_and_functions} 
\end{figure}

First, we introduce different classes of sets and functions. Figure~\ref{fig:diagram_of_subanalytic_sets_and_functions} (Figure 3 in \citet{amir_et_al_finite_witness_thm}) summarizes the relationships. See the appendix of \citet{amir_et_al_finite_witness_thm} for more details.

\begin{definition}[Definition 1.1.1 in \citep{Valette_subanalytic}]
	A subset $E\subseteq \mathbb{R}^n$ is called \textit{semianalytic} if it is locally defined
	by finitely many real analytic equalities and inequalities. Namely, for each $a\in \mathbb{R}^n$,
	there is a neighborhood $U$ of $a$, and real analytic functions $f_{ij}, g_{ij}$ on $U$, where
	$i=1,\ldots,r$ and $ j= 1,\ldots s_i$, such that
	\begin{equation}\label{eq:semi}
	E\cap U=\bigcup_{i=1}^r \bigcap_{j=1}^{s_i}\{\mathbf{z}\in U| \, g_{ij}(\mathbf{z})>0 \text{ and } f_{ij}(\mathbf{z})=0 \}. \end{equation}
\end{definition}

\begin{definition}[Definition 1.1.3 in \citep{Valette_subanalytic}]
	A subset $Z\subseteq \mathbb{R}^n$ is \textit{globally semianalytic} if $V_n(Z)$ is a semianalytic
	subset of $\mathbb{R}^n$, where $V_n : \mathbb{R}^n \to (-1,1)^n$ is the homeomorphism defined by
	$$V_n\left(\mathbf{z}=(z_1,\ldots,z_n)\right)=\left(\frac{z_1}{\sqrt{1+|\mathbf{z}|^2}},\ldots , \frac{z_n}{\sqrt{1+|\mathbf{z}|^2}} \right). $$
\end{definition}

\begin{definition}[Definition 1.1.6 in \citep{Valette_subanalytic}]
	A subset $E\subseteq \mathbb{R}^n$ is \textit{globally subanalytic} if it can be presented
	as a linear projection of a globally semianalytic set; more precisely, if there
	exists a globally semianalytic set $Z \subseteq \mathbb{R}^{n+p}$  such that $E=\pi(Z) $, with   $\pi:\mathbb{R}^{n+p}\to \mathbb{R}^n$ being the projection operator that omits the last $p$ coordinates while leaving the remaining coordinates unchanged.
\end{definition}

\begin{definition}[Definition A.10 in \citep{amir_et_al_finite_witness_thm}]\label{def:sigma_subanalytic_set}
We say that a subset $A\subseteq\mathbb{R}^D$ is $\sigma$\textit{-subanalytic} if it is a countable union of globally subanalytic subsets of $\mathbb{R}^D$.
\end{definition}

\begin{definition}[Definition A.13 in \citep{amir_et_al_finite_witness_thm}]\label{def:sigma_subanalytic_function}
Let $\mathbb{M}\subseteq \mathbb{R}^D$ be a $\sigma$-subanalytic set. We say that $f:\mathbb{M}\xrightarrow[]{}\mathbb{R}^L$ is a $\sigma$\textit{-subanalytic function} if its graph is a $\sigma$-subanalytic subset of $\mathbb{R}^{D+L}$.
\end{definition}

Second, we state useful properties of $\sigma$-subanalytic sets.

\begin{theorem}[Proposition A.11 in \citep{amir_et_al_finite_witness_thm}]\label{thm:sigma_subanalytic_properties}
	Assume that $A,B \subseteq \mathbb{R}^D $ and $C \subseteq \mathbb{R}^M$ are $\sigma$-subanalytic sets. Then:
	\begin{enumerate}
		\item $A \cup B$ is $\sigma$-subanalytic. More generally, any countable union of $\sigma$-subanalytic sets is $\sigma$-subanalytic.
        \item $A\cap B$ is $\sigma$-subanalytic.
		\item $A\times C $ is $\sigma$-subanalytic.
		\item If $\pi:\mathbb{R}^D\to \mathbb{R}^L $ is a linear projection, then $\pi(A) $ is a $\sigma$-subanalytic set.
		\item $A $ is a \emph{countable} union of $C^\infty $ manifolds.
	\end{enumerate}
\end{theorem}

Third, we present a corollary of the finite witness theorem (Theorem A.2) from \citet{amir_et_al_finite_witness_thm}. The Hausdorff dimension is the notion of dimension used. The result holds for Lebesgue-almost every $\left( \vartheta_{1},\ldots,\vartheta_{2D+1}\right)$ when $\mathbb{W}=\mathbb{R}^q$ or is an open subset of $\mathbb{R}^q$.

\begin{theorem}[Corollary A.20 in \citep{amir_et_al_finite_witness_thm}]\label{thm:finite_witness}
Let $\mathbb{M} \subseteq \mathbb{R}^p$, $\mathbb{W} \subseteq \mathbb{R}^q$ be $\sigma$-subanalytic sets of dimension $D$ and $D_{\vartheta}$ respectively. Let $F:\mathbb{M} \times \mathbb{W} \to \mathbb{R}$ be a $\sigma$-subanalytic function. Define the set
$$\mathcal{N}=\{\left(\mathbf{x},\mathbf{y}\right) \in \mathbb{M} \times \mathbb{M} \mid F(\mathbf{x};\vartheta)=F(\mathbf{y};\vartheta),\ \forall \vartheta\in \mathbb{W} \}.$$
Suppose that for all $\left(\mathbf{x},\mathbf{y}\right) \in \mathbb{M} \times \mathbb{M} \setminus \mathcal{N}$
\begin{equation}\label{cond_dimension_defficiency_for_separation}
\dim\{\vartheta\in \mathbb{W}\mid F(\mathbf{x};\vartheta)=F(\mathbf{y};\vartheta) \}\leq D_{\vartheta}-1.\end{equation}
Then for generic $\left( \vartheta_{1},\ldots,\vartheta_{2D+1}\right) \in \mathbb{W}^{2D+1}$,
\begin{equation}\label{eqn:equality_of_functions_2D+1}
\mathcal{N}=\{\left(\mathbf{x},\mathbf{y}\right) \in \mathbb{M} \times \mathbb{M} \mid F(\mathbf{x};\vartheta_{i})=F(\mathbf{y};\vartheta_{i}),\ \forall i=1,\ldots 2D+1\} .
\end{equation}
Moreover, if $\mathbb{W}$ is an open and connected subset of $\mathbb{R}^q$, and $F(\mathbf{x};\vartheta)$ is analytic as a function of $\vartheta$ for all fixed $\mathbf{x} \in \mathbb{M}$, then the condition from Eq.~\eqref{cond_dimension_defficiency_for_separation} is not required, as it is automatically satisfied.
\end{theorem}

The next result is the measurable maximum theorem from \citet{aliprantis_border_inf_dim_analysis_2006}.

\begin{theorem}[Theorem 18.19 in \citep{aliprantis_border_inf_dim_analysis_2006}]\label{thm:mble_max_theorem}
Let $X$ be a separable metrizable space and $(S,\Sigma)$ a measurable space. Let
$\varphi:S\xrightarrow[]{} X$ be a weakly measurable correspondence with nonempty
compact values, and suppose $f:S\times X\to\mathbb{R}$ is a Carath\'eodory function.
Define the value function $m:S\to\mathbb{R}$ by
$$m(s)=\max_{x\in\varphi(s)} f(s,x),$$
and the correspondence $\mu:S\xrightarrow[]{} X$ of maximizers by $$\mu(s)=\{x\in\varphi(s):f(s,x)=m(s)\}.$$
Then:
\begin{enumerate}
    \item The value function $m$ is measurable.
    \item The argmax correspondence $\mu$ has nonempty and compact values.
    \item The argmax correspondence $\mu$ is measurable and admits a measurable selector.
\end{enumerate}

\end{theorem}

The following result is the Tarski--Seidenberg theorem.
\begin{theorem}[Theorem 1.5 in \citep{bierstone_milman_subanalytic_set}]\label{thm:tarski_seidenberg_theorem}
    The image of a semialgebraic set $X\subset \mathbb{R}^{n+1}$ by the projection $\pi:\mathbb{R}^{n+1}\xrightarrow[]{}\mathbb{R}^n$ is semialgebraic.
\end{theorem}

The next result is the uniqueness theorem for characteristic functions.

\begin{theorem}[Proposition 2.9 in \citep{multivariate_stats_book}]\label{thm:uniqueness_characteristic_function} The characteristic function $c:\mathbb{R}^n \to \mathbb{C}$ of $\mathbf{x}$, defined by
$c(\mathbf{t})
=
c_{\mathbf{x}}(\mathbf{t})
=
\mathbb{E}(e^{i\mathbf{t}'\mathbf{x}})$,
is unique: $$\mathbf{x} \overset{d}{=} \mathbf{y}
\quad\iff\quad
c_{\mathbf{x}}(\mathbf{t})
=
c_{\mathbf{y}}(\mathbf{t}),
\quad
\forall\,\mathbf{t}\in\mathbb{R}^n.$$
\end{theorem}

%\subsection{Proof of Theorem~\ref{thm:fractal_whitney_embed_prev_thm}}\label{subsection:proof_of_fractal_whitney_embed_prev_thm}

% See the proof of Theorem 2.3 in~\citet{embedology_Sauer_et_al_1991}.

%\hfill$\square$

\subsection{Proof of Theorem~\ref{thm:injective_discrete_time}}\label{subsection:proof_of_injective_discrete_time}

The proof is a streamlined version of the proof of Theorem 1 in the companion paper~\citep{random_features_estimation_paper}.

\paragraph*{Step 1: Limiting expectations.} We claim that, for all $\theta \in \Theta$, the limiting expectation values of the $k$ random Fourier features are given by \begin{equation*}    \Phi(\theta)=\int_{\mathbb{R}^{(m+1)\times d}}\varphi(x) \, dP_{\theta}(x).\end{equation*} Observe that, for all $\theta \in \Theta$ and all $r \in \{0,1,\ldots,s\}$, as $n\xrightarrow[]{}\infty$, we have
\begin{equation}\label{eqn:convergence_to_Phi_theta_discrete_time}\left\|\frac{1}{n-m}\sum_{t=m+1}^n \mathbb{E}_{\theta}\left[ \varphi\left(\left[X_{t-m}^{(r)}(\theta),\ldots,X_t^{(r)}(\theta)\right]^{\top}\right)\right] - \Phi(\theta)\right\|=o(1),\end{equation} by upper bounding the $\ell_2$ norm $\left\|\cdot\right\|$ by the $\ell_1$ norm (i.e.\ over the $k$ random features) using the monotonicity of norms, linearity, and applying the weak convergence from Assumption~\ref{asmpt:weak_convergence_of_time_avg_of_distributions_discrete_time}.

\paragraph*{Step 2: Injectivity.} The main ingredient is the finite witness theorem (Theorem~\ref{thm:finite_witness}), so we show that its assumptions are satisfied. Denote $$\mathbb{W}=\mathbb{R}^{d(m+1)}\times (-\pi,\pi).$$

By Assumption~\ref{asmpt:statistical_model_discrete_time}, $\Theta$ is a compact subset of $\mathbb{R}^p$ with non-empty interior, so the Hausdorff dimension is $\mathrm{dim}_H(\Theta)=p$. By Assumption~\ref{asmpt:statistical_model_discrete_time}, $\Theta$ is a semialgebraic subset of $\mathbb{R}^p$, hence it is $\sigma$-subanalytic; see Figure~\ref{fig:diagram_of_subanalytic_sets_and_functions}.

By construction, the random parameters $\vartheta_i$, $i=1,\ldots,k$ from Eq.~\eqref{eqn:parameters_for_the_random_features} for the random Fourier features $\varphi_i$, $i=1,\ldots,k$ from Eq.~\eqref{eqn:random_Fourier_features_order_m} are each random variables taking values in $\mathbb{W}$, which is semialgebraic, so, as explained in Figure~\ref{fig:diagram_of_subanalytic_sets_and_functions}, it is $\sigma$-subanalytic.

By Assumption~\ref{asmpt:statistical_model_discrete_time}, the real and imaginary parts of the map $(\theta,\omega)\mapsto \psi_{\theta}(\omega)$ admit jointly real analytic extensions to $O \times \mathbb{R}^{(m+1)\times d}$, so each $(\theta,\vartheta_i)\mapsto \int_{\mathbb{R}^{(m+1)\times d}}\varphi_i(x) \, dP_{\theta}(x)$, $i=1,\ldots,k$, admits a jointly real analytic extension to $O \times \mathbb{W}$ by using the definition of the characteristic function and each random Fourier feature. Note that the graph of this extension is $\sigma$-subanalytic, and that $\Theta$ is a $\sigma$-subanalytic set as discussed previously. Therefore, since $\sigma$-subanalytic sets are closed under finite intersections and Cartesian products by Theorem~\ref{thm:sigma_subanalytic_properties}, we have that the intersection of the graph of this extension with $\Theta \times \mathbb{W}\times \mathbb{R}$ is $\sigma$-subanalytic.

$\mathbb{W}$ is open and connected, and $(\theta,\vartheta_i)\mapsto \int_{\mathbb{R}^{(m+1)\times d}}\varphi_i(x) \, dP_{\theta}(x)$ is analytic as a function of $\vartheta_i$ for all fixed $\theta\in\Theta$. Hence, the condition from Eq.~\eqref{cond_dimension_defficiency_for_separation} from the finite witness theorem (Theorem~\ref{thm:finite_witness}) is satisfied.

Let $x_1,\ldots,x_{m+1}$ be vectors in $\mathbb{R}^d$, and define $x=(x_1,\ldots,x_{m+1})^{\top}\in\mathbb{R}^{(m+1)\times d}$. Letting $a=(a_1^{\top},\ldots,a_{m+1}^{\top})^{\top}\in\mathbb{R}^{d(m+1)}$, $b\in (-\pi,\pi)$, and $c=(a,b)\in\mathbb{W}$, define $$\phi_c(x)=\cos\left(\sum_{j=1}^{m+1} a_j \cdot x_{j} + b\right).$$ Define $$\mathcal{N}=\left\{\left(\theta_1,\theta_2\right) \in \Theta^2 \mid \int_{\mathbb{R}^{(m+1)\times d}}\phi_c(x) \, dP_{\theta_1}(x) =\int_{\mathbb{R}^{(m+1)\times d}}\phi_c(x) \, dP_{\theta_2}(x),\ \forall c\in\mathbb{W} \right\}.$$ By the finite witness theorem (Theorem~\ref{thm:finite_witness}), for generic parameters $\left( \vartheta_1,\ldots,\vartheta_{k}\right) \in \mathbb{W}^k$ of the $k=2p+1$ random Fourier features $\left( \varphi_1,\ldots,\varphi_{k}\right)$, we have the set equality $$\mathcal{N}=\left\{\left(\theta_1,\theta_2\right) \in \Theta^2 \mid \int_{\mathbb{R}^{(m+1)\times d}}\varphi_i(x) \, dP_{\theta_1}(x) =\int_{\mathbb{R}^{(m+1)\times d}}\varphi_i(x) \, dP_{\theta_2}(x),\ \forall i=1,\ldots,k \right\}.$$ By the discussion directly above Theorem~\ref{thm:finite_witness}, generic here means $\Pi$-almost every $\vartheta$, where $\Pi$ is the joint distribution from Eq.~\eqref{eqn:parameters_for_the_random_features} which is absolutely continuous w.r.t.\ the Lebesgue measure.

Denoting the $\mathbb{R}^{(m+1)\times d}$-valued random matrix by $\Omega_i=(\Omega_{i,1},\ldots,\Omega_{i,m+1})^{\top}$, we have \begin{align*}&\int_{\mathbb{R}^{(m+1)\times d}}\varphi_i(x) \, dP_{\theta}(x)
\\& 
=\int_{\mathbb{R}^{(m+1)\times d}}\cos\left(\sum_{j=1}^{m+1} \Omega_{i,j} \cdot x_{j} + \alpha_i\right)\, dP_{\theta}(x)
\\&
=
\cos(\alpha_i)
\int_{\mathbb{R}^{(m+1)\times d}}\cos\left(\sum_{j=1}^{m+1} \Omega_{i,j} \cdot x_{j}\right)\, dP_{\theta}(x) 
\\&
- 
\sin(\alpha_i)
\int_{\mathbb{R}^{(m+1)\times d}}\sin\left(\sum_{j=1}^{m+1} \Omega_{i,j} \cdot x_{j}\right)\, dP_{\theta}(x) 
\\&
= \cos(\alpha_i) \  \mathrm{Re}( \psi_{\theta}(\Omega_i)) - \sin(\alpha_i) \ \mathrm{Im}(  \psi_{\theta}(\Omega_i))
\\&
= M_{\vartheta_i}(\theta),
\end{align*}
where $\psi_{\theta}$ is the characteristic function. If $(\theta_1,\theta_2)\in\mathcal{N}$, then $ M_{\vartheta_i}(\theta_1)=M_{\vartheta_i}(\theta_2)$ for every possible value of $\vartheta_i=(\mathrm{Vec}(\Omega_i),\alpha_i)\in\mathbb{W}$. In particular, the expectation values of the random Fourier features correspond to the real and imaginary parts of the characteristic function by taking $\alpha_i=0$ and $\alpha_i=-\pi/2$, respectively, so by the uniqueness theorem (Theorem~\ref{thm:uniqueness_characteristic_function}) the corresponding distributions must be equal, $P_{\theta_1}=P_{\theta_2}$, if $(\theta_1,\theta_2)\in\mathcal{N}$. However, by Assumption~\ref{asmpt:statistical_model_discrete_time}, the map $\theta\mapsto P_{\theta}$ is injective, i.e., if $\theta_1 \neq \theta_2$ then $P_{\theta_1}\neq P_{\theta_2}$, so we must have
$$(\theta_1,\theta_2)\in\mathcal{N}\implies (\theta_1,\theta_2)\in \{(\theta_1,\theta_2)\in \Theta^2 : \theta_1=\theta_2\}=\{(\theta,\theta):\theta\in\Theta\},$$ so $\mathcal{N}\subseteq \{(\theta,\theta):\theta\in\Theta\}$. Conversely, we have $\{(\theta,\theta):\theta\in\Theta\} \subseteq \mathcal{N}$ trivially. Therefore,
\begin{equation}\label{eqn:conclusion_mathcal_N_discrete_time}\mathcal{N}=\{(\theta,\theta):\theta\in\Theta\}.\end{equation}

Putting it all together, for $\Pi$-almost any random parameters $(\vartheta_1,\ldots,\vartheta_k)$ for the $k=2p+1$ random Fourier features $(\varphi_1,\ldots,\varphi_k)$, for every pair of parameter values $\theta_1, \theta_2\in\Theta$, if $\theta_1\neq \theta_2$, then the expectations of the random Fourier features are not equal, $\Phi(\theta_1)\neq \Phi(\theta_2)$. Precisely in this sense, the map $\theta\mapsto\Phi(\theta)$ is one-to-one for generic choices of $k=2p+1$ random Fourier features.

\hfill$\square$

\subsection{Proof of Theorem~\ref{thm:injective_continuous_time}}\label{subsection:proof_of_injective_continuous_time}

The proof is a streamlined version of the proof of Theorem 3 in the companion paper~\citep{random_features_estimation_paper}.

\paragraph*{Step 1: Local expectations.} We claim that $\forall \tilde{\theta}_u\in\tilde{\Theta}_u$ the local expectation values of the random Fourier features are given by \begin{equation*}   \tilde{\Phi}_u(\tilde{\theta}_u)=\int_{\mathbb{R}^{(m+1)\times d}}\varphi(x) \, d\tilde{P}_{\tilde{\theta}_u,u}(x), \quad u\in [0,1].\end{equation*}  Observe that, for all $\theta \in \Theta$, all $u\in [0,1]$, and all $r \in \{0,1,\ldots,s\}$, we have
\begin{equation*}\begin{aligned}\label{eqn:convergence_to_tilde_Phi_theta_u_continuous_time} \tilde{\Phi}_u(\tilde{\theta}_u)&=\mathbb{E}_{\tilde{\theta}_u}\left[ \varphi\left(\left[\widetilde{X}_{t-m}'^{(r)}(u,\tilde{\theta}_u),\ldots,\widetilde{X}_{t}'^{(r)}(u,\tilde{\theta}_u)\right]^{\top}\right)\right]\\& =\mathbb{E}_{\theta}\left[ \varphi\left(\left[\widetilde{X}_{t-m}^{(r)}(u,\theta),\ldots,\widetilde{X}_{t}^{(r)}(u,\theta)\right]^{\top}\right)\right],\end{aligned}\end{equation*} by using the equivalent representation from Assumption~\ref{asmpt:reparametrization_continuous_time}.

\paragraph*{Step 2: Injectivity.} The two key steps are: (1) the use of the finite witness theorem (Theorem~\ref{thm:finite_witness}), and (2) the extension to almost every rescaled time $u\in [0,1]$. Denote $$\mathbb{W}=\mathbb{R}^{d(m+1)}\times (-\pi,\pi).$$ 

\textbf{Step 2.1.} We first show that the assumptions of Theorem~\ref{thm:finite_witness} are satisfied at each fixed $u$.

For each $u \in [0,1]$, the Hausdorff dimension of $\tilde{\Theta}_u$ is less than or equal to $p$ because (1)  $\Theta$ has Hausdorff dimension $p$ as it is a compact subset of $\mathbb{R}^p$ with non-empty interior by Assumption~\ref{asmpt:statistical_model_continuous_time}, and (2) $\tilde{\Theta}_u=g_u(\Theta)$ and $g_u$ is Lipschitz by Assumption~\ref{asmpt:reparametrization_continuous_time} so $\mathrm{dim}_H(\tilde{\Theta}_u)=\mathrm{dim}_H(g_u(\Theta))\leq \mathrm{dim}_H(\Theta)=p$. Hence, for each $u\in [0,1]$, its Hausdorff dimension is $\mathrm{dim}_H(\tilde{\Theta}_u)\leq p$. Therefore, it suffices to use $k=2p+1$ random Fourier features with Theorem~\ref{thm:finite_witness} (any $k\geq 2p+1$ works, see the proof of Proposition 3.6 in \citep{amir_et_al_finite_witness_thm}).

%%% Hausdorff dim <= box counting dim (when it exists, of course), see page 15 of https://pi.math.cornell.edu/~fractals/6/slides/Lapidus.pdf

By Assumption~\ref{asmpt:statistical_model_continuous_time}, $\Theta$ is a semialgebraic subset of $\mathbb{R}^p$. By Assumption~\ref{asmpt:reparametrization_continuous_time}, $g_u$ is a semialgebraic map. Therefore, by the Tarski--Seidenberg theorem (Theorem~\ref{thm:tarski_seidenberg_theorem}), $\tilde{\Theta}_u=g_u(\Theta)$ is a semialgebraic set (i.e., by using the graph of $g_u$ and repeatedly applying Theorem~\ref{thm:tarski_seidenberg_theorem}), hence it is $\sigma$-subanalytic; see Figure~\ref{fig:diagram_of_subanalytic_sets_and_functions}.

By construction, the random parameters $\vartheta_i$, $i=1,\ldots,k$ from Eq.~\eqref{eqn:parameters_for_the_random_features} for the random Fourier features $\varphi_i$, $i=1,\ldots,k$ from Eq.~\eqref{eqn:random_Fourier_features_order_m} are each random variables taking values in $\mathbb{W}$, which is semialgebraic and hence $\sigma$-subanalytic.

By Assumption~\ref{asmpt:statistical_model_continuous_time}, the real and imaginary parts of the map $(\tilde{\theta}_u,\omega)\mapsto \tilde{\psi}_{\tilde{\theta}_u,u}(\omega)$ admit jointly real analytic extensions to $\tilde{O}_u \times \mathbb{R}^{(m+1)\times d}$, so each 
$(\tilde{\theta}_u,\vartheta_i)\mapsto \int_{\mathbb{R}^{(m+1)\times d}}\varphi_i(x) \, d\tilde{P}_{\tilde{\theta}_u,u}(x)$, $i=1,\ldots,k$, admits a jointly real analytic extension to $\tilde{O}_u\times \mathbb{W}$ by using the definition of the characteristic function and each random Fourier feature. Note that the graph of this extension is $\sigma$-subanalytic, and that $\tilde{\Theta}_u$ is a $\sigma$-subanalytic set as discussed previously. Therefore, since $\sigma$-subanalytic sets are closed under finite intersections and Cartesian products by Theorem~\ref{thm:sigma_subanalytic_properties}, we have that the intersection of the graph of this extension with $\tilde{\Theta}_u \times \mathbb{W}\times\mathbb{R}$ is $\sigma$-subanalytic.

$\mathbb{W}$ is open and connected, and $(\tilde{\theta}_u,\vartheta_i)\mapsto \int_{\mathbb{R}^{(m+1)\times d}}\varphi_i(x) \, d\tilde{P}_{\tilde{\theta}_u,u}(x)$ is analytic as a function of $\vartheta_i$ for all fixed $\tilde{\theta}_u\in\tilde{\Theta}_u$. Hence, the condition from Eq.~\eqref{cond_dimension_defficiency_for_separation} from the finite witness theorem (Theorem~\ref{thm:finite_witness}) is satisfied.

Let $x_1,\ldots,x_{m+1}$ be vectors in $\mathbb{R}^d$, and define $x=(x_1,\ldots,x_{m+1})^{\top}\in\mathbb{R}^{(m+1)\times d}$. Letting $a=(a_1^{\top},\ldots,a_{m+1}^{\top})^{\top}\in\mathbb{R}^{d(m+1)}$, $b\in (-\pi,\pi)$, and $c=(a,b)\in\mathbb{W}$, define $$\phi_c(x)=\cos\left(\sum_{j=1}^{m+1} a_j \cdot x_{j} + b\right).$$ For each fixed $u\in [0,1]$, define $$\mathcal{N}=\left\{(\tilde{\theta}_{u,1},\tilde{\theta}_{u,2}) \in \tilde{\Theta}_u^2 \mid \int_{\mathbb{R}^{(m+1)\times d}}\phi_c(x)  d\tilde{P}_{\tilde{\theta}_{u,1},u}(x) =\int_{\mathbb{R}^{(m+1)\times d}}\phi_c(x) d\tilde{P}_{\tilde{\theta}_{u,2},u}(x), \forall c\in\mathbb{W} \right\},$$ where we have suppressed the dependence of $\mathcal{N}$ on $u$. For each fixed $u\in [0,1]$, by the finite witness theorem (Theorem~\ref{thm:finite_witness}), for generic parameters $\left( \vartheta_1,\ldots,\vartheta_{k}\right) \in \mathbb{W}^k$ of the $k=2p+1$ random Fourier features $\left( \varphi_1,\ldots,\varphi_{k}\right)$, we have the set equality $$\mathcal{N}=\left\{(\tilde{\theta}_{u,1},\tilde{\theta}_{u,2}) \in \tilde{\Theta}_u^2 \mid \int_{\mathbb{R}^{(m+1)\times d}}\varphi_i(x) d\tilde{P}_{\tilde{\theta}_{u,1},u}(x) =\int_{\mathbb{R}^{(m+1)\times d}}\varphi_i(x) d\tilde{P}_{\tilde{\theta}_{u,2},u}(x),  \forall i\in [k]\right\}.$$  By the discussion above Theorem~\ref{thm:finite_witness}, generic here means $\Pi$-almost every $\vartheta$, where $\Pi$ is the joint distribution from Eq.~\eqref{eqn:parameters_for_the_random_features} which is absolutely continuous w.r.t.\ the Lebesgue measure.

Denoting the $\mathbb{R}^{(m+1)\times d}$-valued random matrix by $\Omega_i=(\Omega_{i,1},\ldots,\Omega_{i,m+1})^{\top}$, we have \begin{align*}&\int_{\mathbb{R}^{(m+1)\times d}}\varphi_i(x) \, d\tilde{P}_{\tilde{\theta}_u,u}(x)
\\& 
=\int_{\mathbb{R}^{(m+1)\times d}}\cos\left(\sum_{j=1}^{m+1} \Omega_{i,j} \cdot x_{j} + \alpha_i\right)\, d\tilde{P}_{\tilde{\theta}_u,u}(x)
\\&
=
\cos(\alpha_i)
\int_{\mathbb{R}^{(m+1)\times d}}\cos\left(\sum_{j=1}^{m+1} \Omega_{i,j} \cdot x_{j}\right)\, d\tilde{P}_{\tilde{\theta}_u,u}(x) 
\\&
- 
\sin(\alpha_i)
\int_{\mathbb{R}^{(m+1)\times d}}\sin\left(\sum_{j=1}^{m+1} \Omega_{i,j} \cdot x_{j}\right)\, d\tilde{P}_{\tilde{\theta}_u,u}(x) 
\\&
= \cos(\alpha_i) \  \mathrm{Re}( \tilde{\psi}_{\tilde{\theta}_u,u}(\Omega_i)) - \sin(\alpha_i) \ \mathrm{Im}(  \tilde{\psi}_{\tilde{\theta}_u,u}(\Omega_i))
\\&
= M_{\vartheta_i}(\tilde{\theta}_u,u),
\end{align*}
where $\tilde{\psi}_{\tilde{\theta}_u,u}$ is the characteristic function. If $(\tilde{\theta}_{u,1},\tilde{\theta}_{u,2})\in\mathcal{N}$, then $ M_{\vartheta_i}(\tilde{\theta}_{u,1},u)= M_{\vartheta_i}(\tilde{\theta}_{u,2},u)$ for every possible value of $\vartheta_i=(\mathrm{Vec}(\Omega_i),\alpha_i)\in\mathbb{W}$. In particular, the expectation values of the random Fourier features correspond to the real and imaginary parts of the characteristic function by taking $\alpha_i=0$ and $\alpha_i=-\pi/2$, respectively, so by the uniqueness theorem (Theorem~\ref{thm:uniqueness_characteristic_function}) the corresponding distributions must be equal, $\tilde{P}_{\tilde{\theta}_{u,1},u}= \tilde{P}_{\tilde{\theta}_{u,2},u}$, if $(\tilde{\theta}_{u,1},\tilde{\theta}_{u,2})\in\mathcal{N}$. However, by Assumption~\ref{asmpt:statistical_model_continuous_time}, the map $\tilde{\theta}_u\mapsto \tilde{P}_{\tilde{\theta}_u,u}$ is injective, i.e., if $\tilde{\theta}_{u,1} \neq \tilde{\theta}_{u,2}$ then $\tilde{P}_{\tilde{\theta}_{u,1},u}\neq \tilde{P}_{\tilde{\theta}_{u,2},u}$, so we must have $$(\tilde{\theta}_{u,1},\tilde{\theta}_{u,2})\in\mathcal{N}\implies (\tilde{\theta}_{u,1},\tilde{\theta}_{u,2})\in \{(\tilde{\theta}_{u,1},\tilde{\theta}_{u,2})\in\tilde{\Theta}_u^2:\tilde{\theta}_{u,1}=\tilde{\theta}_{u,2}\}=\{(\tilde{\theta}_u,\tilde{\theta}_u):\tilde{\theta}_u\in\tilde{\Theta}_u\},$$ so $\mathcal{N}\subseteq \{(\tilde{\theta}_u,\tilde{\theta}_u):\tilde{\theta}_u\in\tilde{\Theta}_u\}$. Conversely, we have $\{(\tilde{\theta}_u,\tilde{\theta}_u):\tilde{\theta}_u\in\tilde{\Theta}_u\} \subseteq \mathcal{N}$ trivially. Therefore,
\begin{equation}\label{eqn:conclusion_mathcal_N_continuous_time}\mathcal{N}=\{(\tilde{\theta}_u,\tilde{\theta}_u):\tilde{\theta}_u\in\tilde{\Theta}_u\}.\end{equation}

Hence, for any particular rescaled time $u\in [0,1]$, for almost any random parameters $(\vartheta_1,\ldots,\vartheta_k)$ for the $k=2p+1$ random Fourier features $(\varphi_1,\ldots,\varphi_k)$, we have that for any two distinct parameter values, $\tilde{\theta}_{u,1} \neq \tilde{\theta}_{u,2}$, the corresponding expectations of the $k=2p+1$ random Fourier features are not equal, $\tilde{\Phi}_u(\tilde{\theta}_{u,1})\neq \tilde{\Phi}_u(\tilde{\theta}_{u,2})$.

\textbf{Step 2.2.} Next, we extend the result to almost every
rescaled time $u\in[0,1]$.

For each $\ell\in\mathbb N$, define \begin{equation}\label{eqn:function_being_maximized}
b_{\ell}(u,\vartheta)
=
\max_{(\tilde{\theta}_{u,1},\tilde{\theta}_{u,2}) \in\tilde{\Theta}_u^2}
-\left(\norm{\tilde{\Phi}_u
(\tilde{\theta}_{u,1})
-
\tilde{\Phi}_u
(\tilde{\theta}_{u,2})} +
\max\left(
\frac1\ell-
\|\tilde{\theta}_{u,1}-\tilde{\theta}_{u,2}\|
, 0 \right)
\right).
\end{equation}

The function $\bar{\psi}$ from Assumption~\ref{asmpt:measurability} induces an extension of $\tilde{\Phi}_u$ to $\mathbb{R}^{\tilde{p}}$ that is jointly Borel measurable and continuous in $\tilde{\theta}_u$. We continue to denote this extension by $\tilde{\Phi}_u$ to simplify the presentation. By Assumptions~\ref{asmpt:statistical_model_continuous_time} and~\ref{asmpt:measurability}, the set-valued map $$
(u,\vartheta)
\mapsto
\tilde{\Theta}_u^2,$$ % vartheta does nothing here, but it's fine
% cartesian product of compact sets is compact, yes
is measurable and has values that are nonempty compact sets, and the function being maximized in Eq.~\eqref{eqn:function_being_maximized},
\begin{equation}\label{eqn:map_param_to_objective}(\tilde{\theta}_{u,1},\tilde{\theta}_{u,2})\mapsto -\left(\norm{\tilde{\Phi}_u
(\tilde{\theta}_{u,1})
-
\tilde{\Phi}_u
(\tilde{\theta}_{u,2})} +
\max\left(
\frac1\ell-
\|\tilde{\theta}_{u,1}-\tilde{\theta}_{u,2}\|
, 0 \right)
\right),\end{equation} is jointly Borel measurable and continuous
in
$(\tilde{\theta}_{u,1},\tilde{\theta}_{u,2})$
for each fixed $(u,\vartheta)$. Hence, by the
measurable maximum theorem (Theorem~\ref{thm:mble_max_theorem}), the function $b_{\ell}$ from Eq.~\eqref{eqn:function_being_maximized} is Borel measurable. Then the set \begin{equation}\label{eqn:set_B}B
=
\bigcup_{\ell=1}^{\infty}
\left\{
(u,\vartheta):
b_{\ell}(u,\vartheta)=0
\right\},\end{equation} is a Borel set. This fact will be used later on, so that we may apply Tonelli's theorem.

Next, we show that $(u,\vartheta)\in B$ if and only if
$\tilde{\Phi}_u$ is not one-to-one. 

First, suppose that $\tilde{\Phi}_u$ is not one-to-one. Then there exist distinct parameters $\tilde{\theta}_{u,1},\tilde{\theta}_{u,2}\in\tilde{\Theta}_u$ and some $\ell\in\mathbb{N}$ such that $\norm{\tilde{\theta}_{u,1}-\tilde{\theta}_{u,2}}\geq \frac{1}{\ell}$ and $\tilde{\Phi}_u
(\tilde{\theta}_{u,1})
=
\tilde{\Phi}_u
(\tilde{\theta}_{u,2})$. Therefore, we have that $b_{\ell}(u,\vartheta)=0$ for some $\ell\in\mathbb{N}$, so $(u,\vartheta)\in B$ by their definitions Eq.~\eqref{eqn:function_being_maximized} and Eq.~\eqref{eqn:set_B}.

Second, conversely, suppose that $(u,\vartheta)\in B$. Then there exists some $\ell\in\mathbb{N}$ such that $b_{\ell}(u,\vartheta)=0$ by definition. By the compactness of $\tilde{\Theta}_u$ from Assumption~\ref{asmpt:statistical_model_continuous_time} and the continuity of the map from Eq.~\eqref{eqn:map_param_to_objective}, the extreme value theorem implies that the map from Eq.~\eqref{eqn:map_param_to_objective} must attain its maximum at some $(\tilde{\theta}_{u,1},\tilde{\theta}_{u,2})\in\tilde{\Theta}_u^2$. Therefore, both terms from Eq.~\eqref{eqn:function_being_maximized}, $$\norm{\tilde{\Phi}_u
(\tilde{\theta}_{u,1})
-
\tilde{\Phi}_u
(\tilde{\theta}_{u,2})},\quad \max\left(
\frac1\ell-
\|\tilde{\theta}_{u,1}-\tilde{\theta}_{u,2}\|
, 0 \right),$$ must be equal to zero for some $\tilde{\theta}_{u,1},\tilde{\theta}_{u,2}\in\tilde{\Theta}_u$. That is, there exist distinct parameter values $\tilde{\theta}_{u,1},\tilde{\theta}_{u,2}\in\tilde{\Theta}_u$ and some $\ell \in \mathbb{N}$ such that $\norm{\tilde{\theta}_{u,1}-\tilde{\theta}_{u,2}}\geq \frac{1}{\ell}$ and $\tilde{\Phi}_u
(\tilde{\theta}_{u,1})
=
\tilde{\Phi}_u
(\tilde{\theta}_{u,2})$. Therefore, $\tilde{\Phi}_u$ is not one-to-one.

Let $\Pi$ denote the joint distribution of the random parameters $\vartheta=(\vartheta_1,\ldots,\vartheta_k)\in
\mathbb{W}^k$ from Eq.~\eqref{eqn:parameters_for_the_random_features}. Note that $\Pi$ is absolutely continuous
with respect to Lebesgue measure on $\mathbb{W}^k$. We just established that $(u,\vartheta)\in B$ if and only if
$\tilde{\Phi}_u$ is not one-to-one. By the conclusion of Step 2.1, namely the set equality from Eq.~\eqref{eqn:conclusion_mathcal_N_continuous_time} due to the finite witness theorem, and the absolute continuity of $\Pi$ w.r.t.\ the Lebesgue measure, for every fixed $u\in [0,1]$, we have $$\Pi\left(\left\{\vartheta\in\mathbb{W}^k:
(u,\vartheta)\in B
\right\}\right) =0.$$ Therefore, we have
$$\int_{[0,1]}\Pi\left(\left\{\vartheta\in\mathbb{W}^k:
(u,\vartheta)\in B
\right\}\right) d\lambda(u)=0.$$

We previously showed that $B$ from Eq.~\eqref{eqn:set_B} is a Borel set, hence the indicator function $$1_B(u,\vartheta)=\begin{cases}
    1, \quad (u,\vartheta)\in B, \\0, \quad (u,\vartheta)\not\in B,
\end{cases}$$ is jointly Borel measurable. Thus, we may apply Tonelli's theorem. Observe that \begin{align*}\label{eqn:conclusion_tonelli}
&\int_{\mathbb{W}^k}
\lambda\left(
\left\{
u\in[0,1]:
(u,\vartheta)\in B
\right\}
\right) d\Pi(\vartheta)\\
&
=\int_{\mathbb{W}^k} \left(\int_{[0,1]}
1_B(u,\vartheta) d\lambda(u)\right)  d\Pi(\vartheta)\\
&
=\int_{[0,1]}\left(\int_{\mathbb{W}^k} 1_B(u,\vartheta) d\Pi(\vartheta) \right) d\lambda(u)\\
&
=
\int_{[0,1]}\Pi\left(\left\{\vartheta\in\mathbb{W}^k:
(u,\vartheta)\in B
\right\}\right) d\lambda(u)
\\&
=
0.
\end{align*} The integrand $\lambda\left(
\left\{
u\in[0,1]:
(u,\vartheta)\in B
\right\}
\right)$ is nonnegative, so for $\Pi$-almost every $\vartheta$, we have  
$$\lambda\left(
\left\{
u\in[0,1]:
(u,\vartheta)\in B
\right\}
\right)=0.$$
However, we previously showed that $(u,\vartheta)\in B$ if and only if
$\tilde{\Phi}_u$ is not one-to-one, so for $\Pi$-almost every $\vartheta$, we have
$$\lambda\left(
\left\{
u\in[0,1]:
\tilde{\Phi}_u \text{ is not one-to-one }
\right\}
\right)=0.$$
Putting it all together, for $\Pi$-almost any random parameters $\vartheta=(\vartheta_1,\ldots,\vartheta_k)$ for the $k=2p+1$ random Fourier features $(\varphi_1,\ldots,\varphi_k)$, there exists a subset of rescaled times with full measure with the property that $\tilde{\Phi}_u$ is one-to-one, i.e., for $\Pi$-almost every $\vartheta$, 
\begin{equation}\label{eqn:set_of_rescaled_times_one_to_one}\exists \ \mathcal{U}_{\vartheta}=\left\{
u\in[0,1]:
\tilde{\Phi}_u \text{ is one-to-one }
\right\}\subseteq [0,1] \text{ such that } \lambda\left(\mathcal{U}_{\vartheta}\right)=1.\end{equation}

\hfill$\square$

\bibliography{identification_paper_ref}% Produces the bibliography via BibTeX.

\end{document}